\documentclass[seceq]{ptptex}

\usepackage{graphicx}

\usepackage{wrapft}

\def\lsim{\mathrel{\rlap{\lower 4pt \hbox{\hskip 1pt $\sim$}}\raise 1pt \hbox
        {$<$}}}
\def\gsim{\mathrel{\rlap{\lower 4pt \hbox{\hskip 1pt $\sim$}}\raise 1pt \hbox
        {$>$}}}

\def\x12c{$X_{{\rm C}}(^{12}$C)}
\def\cag{$^{12}$C($\alpha,\gamma)^{16}$O }
\def\msun{$M_\odot$}

\pubinfo{Vol.~0, No.~0, April 2012}

\title{
Massive Star Evolution and Nucleosynthesis 
}
\subtitle{Lower End of Fe-Core Collapse Supernova Progenitors and 
\\Remnant Neutron Star Mass Distribution}

\author{
Hideyuki \textsc{Umeda},$^{1}$
Takashi \textsc{Yoshida},$^{1}$
and Koh \textsc{Takahashi}$^{1}$
}

\inst{
$^1$
Department of Astronomy, Graduate School of Science,
University of Tokyo, Tokyo 113-0033, Japan
}

\recdate{
April , 2012; revised , 2012}

\abst{
 In order to explore various aspects of stellar evolution, 
supernovae, gamma ray bursts and nucleosynthesis, we have developed 
a new efficient stellar evolution code. In this paper we describe
this new code and compare the results with the ones calculated by the
previous code. Specifically we focus on the progenitor evolution of 
lower end of the Fe-core collapse supernovae, and mass distribution
of remnant neutron stars. We describe how different assumptions will 
lead different neutron star mass distribution.
We also review recent works of our research group.
}

\begin{document}
\maketitle

\section{Introduction}

 Massive stars end their life as supernovae 
leaving neutron stars behind, or by forming 
blackholes without explosions if they are not rotating. 
The critical initial stellar mass for the blackhole formation is usually
considered to be about 20 to 25 $M_\odot$ but it is uncertain. \cite{Smartt09}
This is because
explosion mechanism for supernovae is not yet known definitely, 
\cite{Herant94,Blondin03,Kotake04,Burrows06,Ohnishi06,Takiwaki12}
and also there are still uncertainties in the progenitor models.


 Lower end for the neutron star forming supernova is also still
uncertain (e.g., Ref. \citen{Poelarends08}). 
Here, the uncertainties in the stellar evolution theory
may be even larger. It is well known that the stars above around 
10 $M_\odot$ stars form an Fe core in the end of their evolution.
The formation of
Fe core is relatively simple for a $M > 13 M_\odot$ star (review in Ref. \citen{Woosley02}). 
In such a star the hottest region is the center mostly all the time through the
evolution. Therefore, heaviest element is
synthesized around the center, forming  Fe-core in the end.

 On the other hand the evolution of $M \lsim 13 M_\odot$ stars are 
more complicated.
Less massive stars 
tend to have temperature inversion between the center and outside 
regions. This phenomenon is well known for intermediate mass stars
with $M < 8 M_\odot$ after carbon burning, for which the carbon burning starts
at off center (e.g., Ref.~\citen{HU1999}). 
For a $M \lsim 11 M_\odot$ star, off center O- and/or Si-burning occurs.
The off-center burning front propagates inward through complicated
burning stages, forming an Fe core eventually.
This off center burning becomes sometimes very violent and it could cause
some mass ejection \cite{WW86,NH88} (see also Section 3.1 in this paper). 
However, such calculations 
have not been done recently with updated input physics, 
so it is currently not clear if such mass ejection really occurs. 

 For less massive stars which do not ignite Ne even at off center,
a cool degenerate O-Ne core is first formed. This O-Ne core
grows in mass gradually when the star is in a super AGB phase. 
Depending on the mass loss rate this O-Ne core reaches the critical
mass for core-collapse owing to electron capture.\cite{Miyaji80,Poelarends08}
It is considered that such a star explodes by the neutrino energy transport
mechanism.\cite{Kitaura06}

 Calculations for the progenitors of
such low mass core-collapse supernovae with updated input physics are 
interesting and important.  
However, their evolution can be quite different if the initial mass is only 
slightly
different, say by 0.01-0.1 $M_\odot$ (e.g., Ref.~\citen{Poelarends08}; see also
Sec. 3.1). 
To understand the whole story in this mass rage, therefore, we need
to calculate stellar evolution in very fine mass grids. 
This will be quite time consuming and thus currently well-used
progenitor models in the literature 
(e.g., Refs.~\citen{NH88,WW95,UN02,UN08,CL98,HW01})
do not fully deal with this mass range.

 To tackle this problem, we have developed a stellar evolution code for efficient 
computation. In this paper we describe this new code, 
the Yoshida-Umeda (YU) code \cite{YU11}, and
compare the results with the ones calculated by one of the author, H.U., using
the Umeda-Nomoto (UN) code \cite{UN02,UN08,UN03,UN05}.
We also briefly review some of the recent works using the YU code 
and other works in our research group.

 This paper is organized as follows. In \S 2, we describe the
new stellar evolution code and differences with previous codes.
In \S 3, massive star evolutions calculated with this code 
are compared with the ones with a previous code. In this section
we also describe the progenitor evolution of lower end of Fe-core collapse SNe
in some detail. \S 4 describes nucleosynthetic aspects 
including some reviews of our recent works. In \S 5 remnant 
neutron star masses are given as a function of progenitor mass.
We describe how the different assumption will lead different
neutron star mass distribution. \S 6 reviews other recent works
of our research group and \S 7 gives discussions and Future Prospects.

\section{Stellar evolution  code}

\subsection{Umeda-Nomoto (UN) code}

 Before describing the YU code, we briefly describe the UN code because 
we will compare the results of these codes. This code is mostly based on
Nomoto-Hashimoto (NH) code \cite{NH88}
and Saio-Nomoto-Kato (SNK) code \cite{SNK88}.
In the UN code, input physics such as equation of state (EOS)
are same as the NH code except updates of radiative opacity,
neutrino emissivity and electron capture rates. 

 There are several differences in the NH and SNK codes: 

i) The NH code is a He-star code, which means that it cannot solve stellar 
atmosphere and hydrogen burning phase. Thus it cannot provide stellar radius
correctly. On the other hand, the SNK code solves atmosphere.

ii) Since the SNK code is not designed for calculating later evolutionary
stages of massive stars, it does not deal with nuclear burning after
carbon burning stages. Also because of that, the SNK code does not include
the acceleration term\cite{KW90} and omit the inertial term
in the equation of motion, while the NH code can include it. 
This acceleration term becomes important
for constructing supernova progenitor models just before iron-core collapse.

iii) The treatment of convection is different. Both codes adopt
Schwarzschild criterion for convection but the way of mixing is 
different. The NH code assumes instantaneous mixing in convective
regions, while in the SNK code matter is mixed diffusively using the
formalism of Ref.~\citen{Spruit92} 
taking into account of the semi-convection effects. 
The energy transfer in a convective region is also different.
In the NH code time-dependent mixing length theory\cite{Unno67,Nomoto77} can be included 
for convective energy transfer, though this effect is not so important 
for the massive star evolution forming an Fe-core.

 In the UN code atmosphere is calculated as the SNK code, the acceleration 
term is included and the formalism of Ref.~\citen{Spruit92} 
is adopted for 
convective mixing. Convective energy transfer is treated as same as the
NH code.  

 The UN code differs from the NH and SNK codes in the calculations
of nucleosynthesis and nuclear energy generation. The NH code assumes 
{\it quasi nuclear statistical equilibrium}
during the Si burning while the UN code solves full nuclear reaction
networks below $\log T ({\rm K}) \lsim 9.6$. 
Above that temperature both codes assume 
{\it nuclear statistical equilibrium} (NSE).
In order to calculate nuclear energy generation rates,
in the UN code nuclear reaction networks are solved simultaneously with
Henyey relaxation, while in the NH and SNK codes abundance is fixed during
Henyey relaxation. In this sense the UN code solves abundance implicitly,
while the NH and SNK codes solve explicitly. Solving abundance implicitly is 
the best way to obtain consistency in the energy generation rates and the
abundance evolution. However, this has a disadvantage in efficient calculations
because solving large reaction networks involves time consuming
matrix inversion calculations.

\subsection{Yoshida \&  Umeda (YU) code}
 
 Basic structure of this code is based on the SNK code \cite{SNK88}.
As mentioned above, although the UN code has been successfully used for the 
progenitor calculations, it has an disadvantage in the calculation time.
To finish a calculation from ZAMS to Fe-core collapse it typically
takes few months. Therefore this code is not suitable for a large parameter
search. Hence, we have developed a new more efficient code (YU code).
Main difference of the YU code from the UN code is the calculations of
nucleosynthesis and energy generation.

The YU code solves a full nuclear reaction network 
from hydrogen burning up to $\log_{10} T_{\rm C} \sim 10.0$.
The nuclear reaction network consists of $\sim 300$ species of nuclei
from $n$, $p$ to Br.
NSE is not assumed in calculations.
On the other hand, the nucleosynthesis is solved before Henyey relaxation
in the YU code similar to the SNK code.
Using the abundance of the next step,
the energy generation is calculated during the Henyey relaxation.
Therefore, smaller time step is required.

After carbon burning stage, the time interval is typically determined
to satisfy the conditions $\Delta \log_{10} T/\log_{10} T \le 0.001$ and
$\Delta \log_{10} \rho / \log_{10} \rho \le 0.003$ for each mass coordinate.
This treatment requires more calculation steps but still saves the
calculation time especially in the late phase of the evolution.
With this code, we can calculate one model
typically in $1 - 2$ weeks for massive star evolutions.

 This code at present has not implemented the acceleration term yet.
We will include that term in near future.

\subsection{Mass loss rates}

 In the UN and YU codes,
same mass loss rates are used for the solar metallicity 
before the Wolf-Rayet star phases shown in this paper.
In OB stars, where the surface temperature is larger than
$1.2 \times 10^{4}$ K and the surface H mass fraction is larger than 0.4,
the mass loss rate is adopted from Ref.~\citen{Vink01}. 
In yellow supergiants and red-giant branches, the mass loss rate in
Ref.~\citen{deJager88} 
is used. In the UN code, metallicity dependent factor of 
$(Z/0.02)^{0.5}$ is multiplied to 
the rate for metal poor stars as 
Ref.~\citen{Kudritzki89}. 
The YU code adopts the metallicity dependence in a different manner.
The metallicity dependence in the main-sequence stage is taken from 
Ref.~\citen{Vink01}. 
In the YU code presented in this paper the case A mass loss rate
in the Ref.~\citen{YU11} is used.
In yellow supergiants and red-giant branches, the metallicity dependent
factor $(Z/0.02)^{0.64}$ is multiplied to the mass loss rate.
The power index is the same as that of B supergiants in 
Ref.~\citen{Vink01}. 
See Ref.~\citen{YU11} for the detail.

\begin{figure}[tbp]
\centerline{
\includegraphics[width=8cm,angle=270]{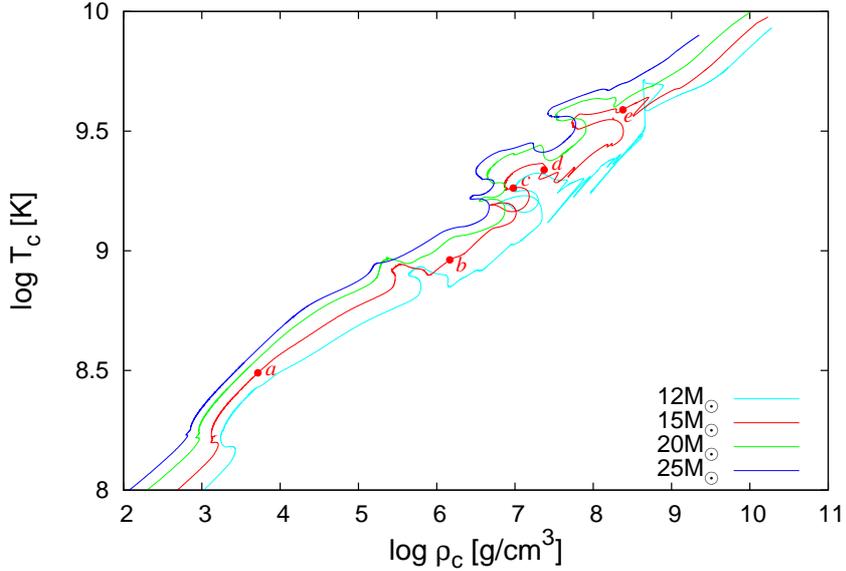}}
\caption{Evolution tracks of $M = 12, 15, 20$, and $25 M_\odot$ stars 
in terms of central temperature and density.
Mass fraction distributions of a $15 M_\odot$ star denoted at the points
$a-e$ are shown in Fig.~\ref{fig_ev15}.
}
\label{fig_KT7}
\end{figure}

\begin{figure}[tbp]
\parbox{\halftext}{
  \centerline{
    \includegraphics[width=4.5cm,angle=270]{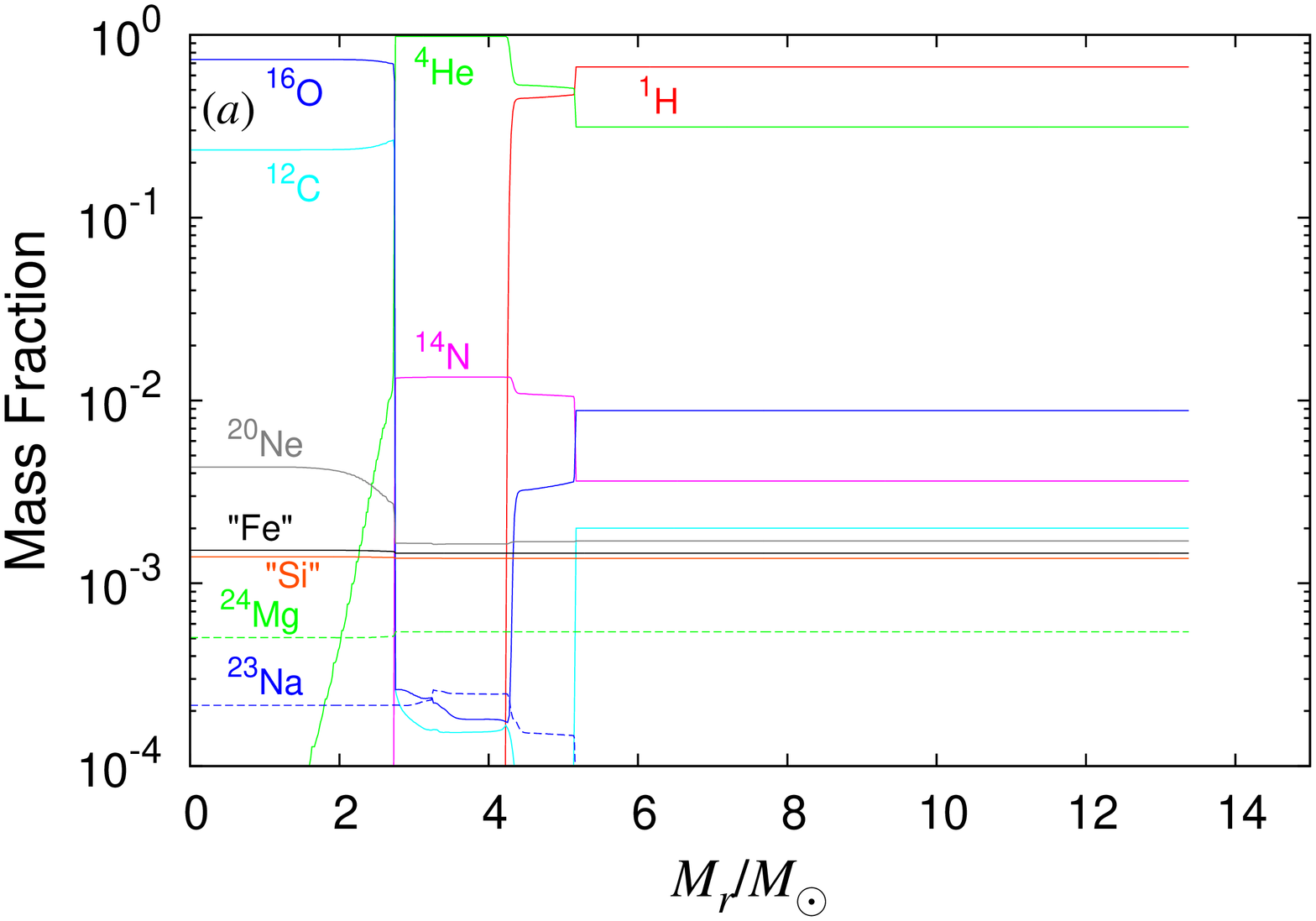}
  }
}
\parbox{\halftext}{
  \centerline{
    \includegraphics[width=4.5cm,angle=270]{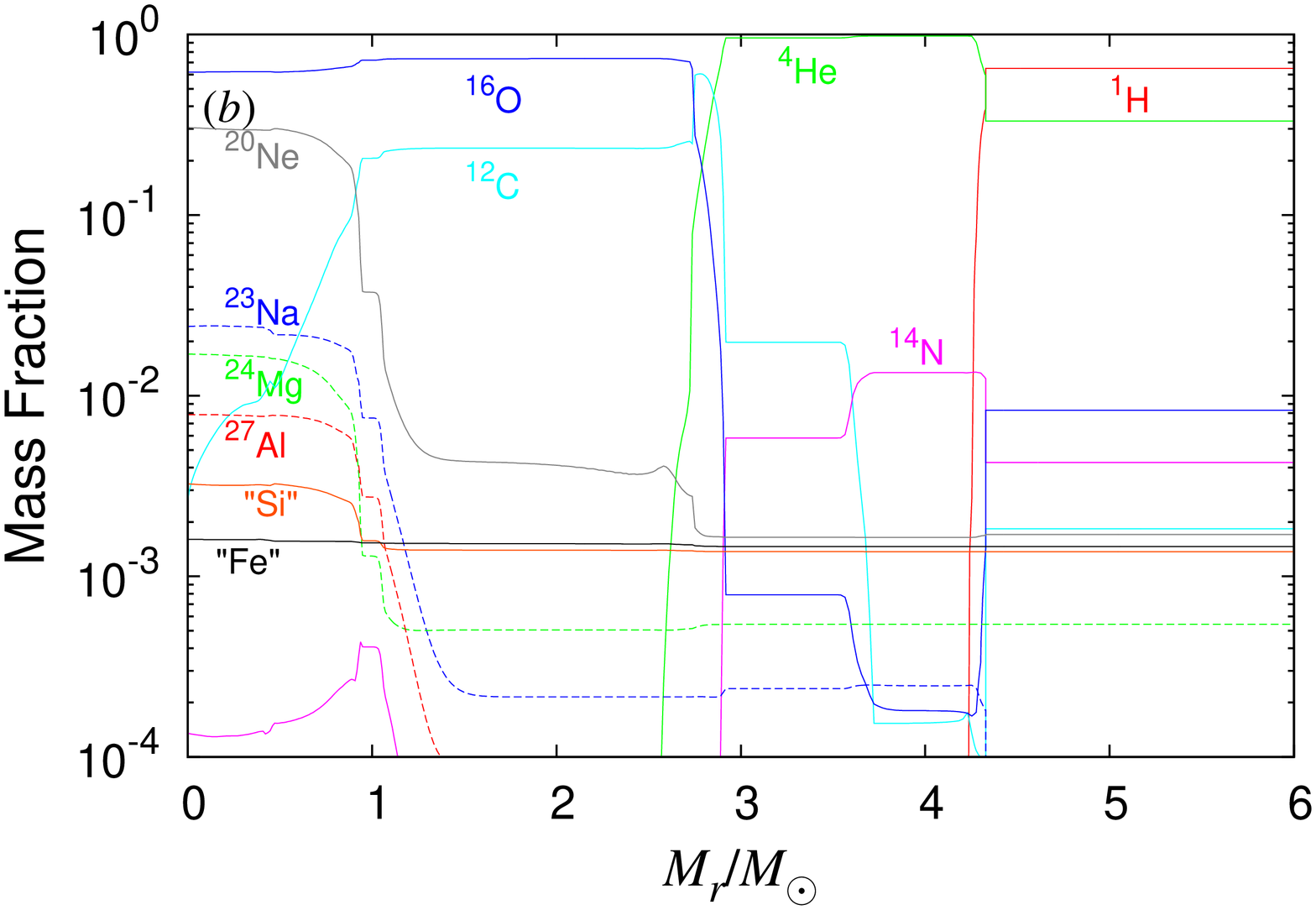}
  }
}

\parbox{\halftext}{
  \centerline{
    \includegraphics[width=4.5cm,angle=270]{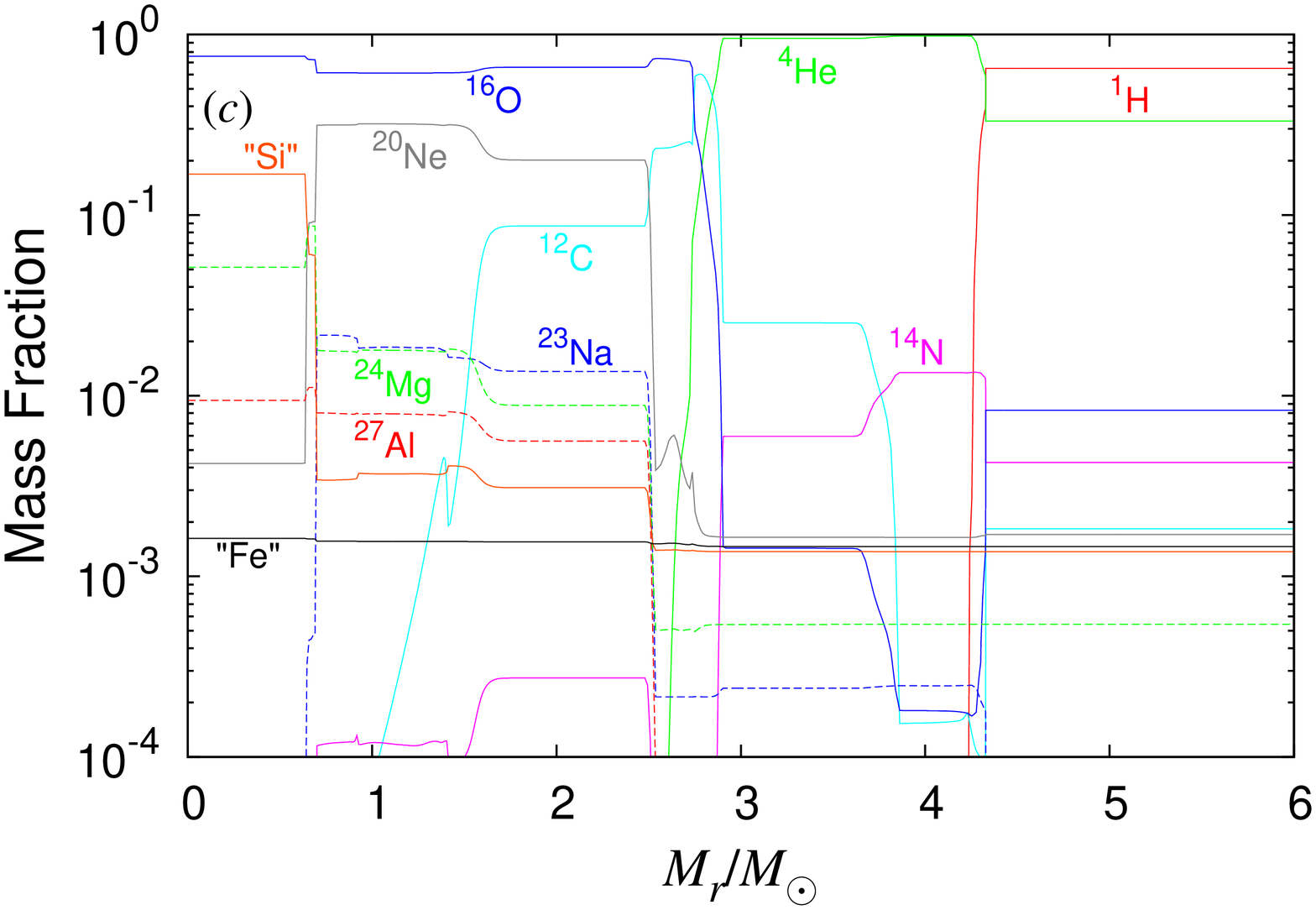}
  }
}
\parbox{\halftext}{
  \centerline{
    \includegraphics[width=4.5cm,angle=270]{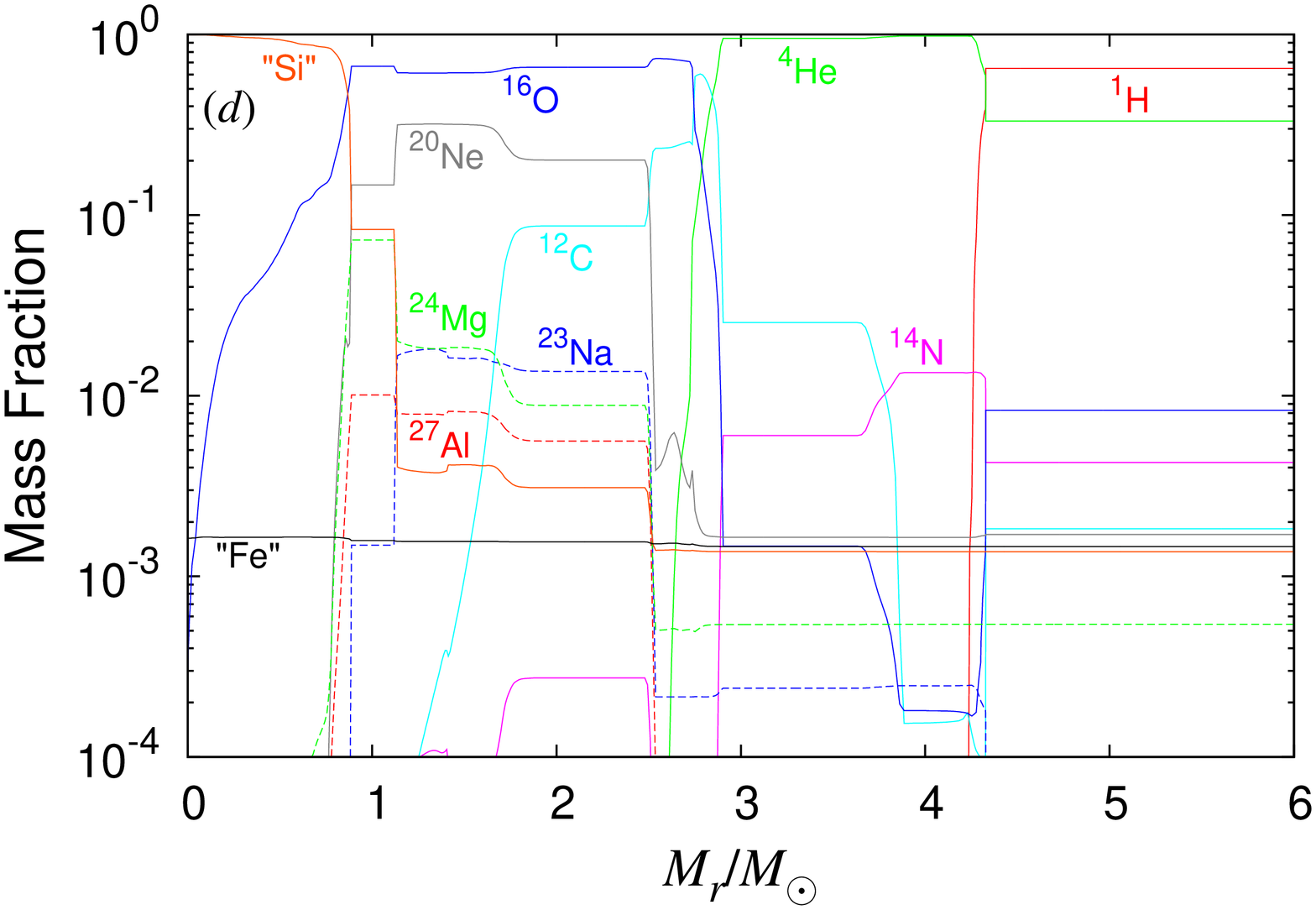}
  }
}

\centerline{
  \parbox{\halftext}{
    \includegraphics[width=4.5cm,angle=270]{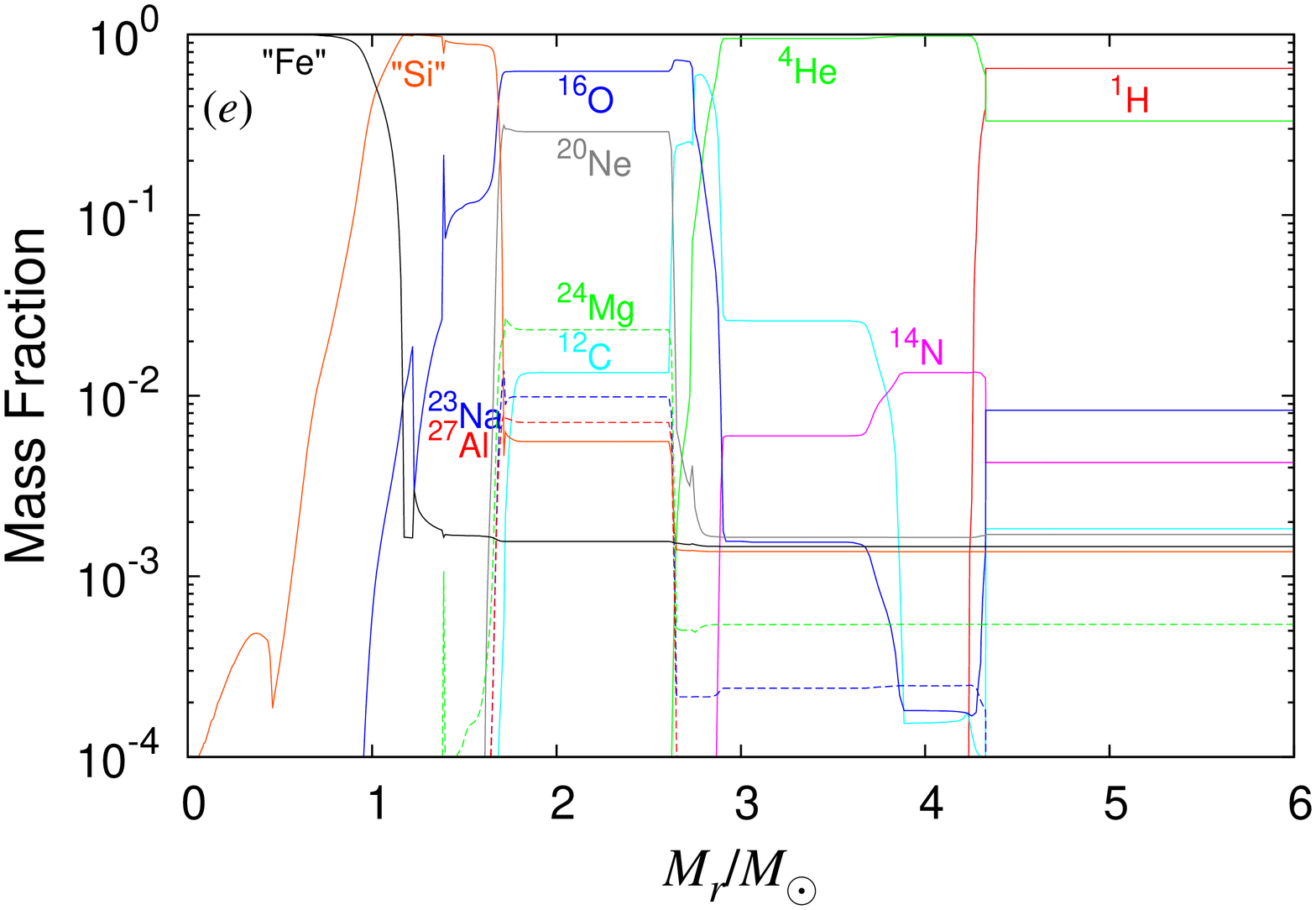}
  }
}
\caption{Mass fraction distributions of a $M = 15 M_\odot$ star
after the He burning ($a$), C burning ($b$), Ne burning ($c$),
O burning ($d$), and Si burning ($e$).
The final mass is $M_{\rm f} = 13.6 M_\odot$.
^^ ^^ Si" and ^^ ^^ Fe" indicate the sum of the mass fractions
of the elements Si-Sc and Ti-Br.
}
\label{fig_ev15}
\end{figure}

\section{Progenitor Evolution}

 Basic properties of massive star evolution have been discussed in many
papers (e.g., Refs.~\citen{NH88,UN02,UN08,CL98,HW01}), 
therefore 
we do not repeat detail here, except $M \simeq 10-13 M_\odot$ models 
in the next subsection.
We briefly describe the evolution in advanced stages of a $M = 15 M_\odot$ 
star as an example.
Fig.~\ref{fig_KT7} shows the evolution tracks of $M = 12, 15, 20$, and $25 M_\odot$
stars in terms of the central density and temperature.
We also show the mass fraction distributions of the 15 $M_\odot$ star at
five stages during the evolution in Fig.~\ref{fig_ev15}.
Fig.~\ref{fig_ev15}($a$) indicates the mass fraction distribution after
the core helium burning.
There is a CO core of 2.2 $M_\odot$ in the central region.
The He layer of 2 $M_\odot$ and the H-rich envelope surround the CO core.
The He layer consists of ashes of the hydrogen burning. 
After the core helium burning, the shell helium burning produces $^{12}$C
in the He layer.
Then, the carbon ignites at the center when the central temperature
becomes $\log_{10} T_{\rm C} ({\rm K}) \sim 8.8$.
The core carbon burning converts $^{12}$C into $^{20}$Ne and $^{23}$Na to
form an O/Ne core (Fig.~\ref{fig_ev15}($b$)).
Some Mg-Si are also produced in the core.

The shell carbon burning is followed by the core carbon burning.
It converts the O/C layer into an O/Ne layer.
However, $^{12}$C of 0.09 by mass fraction remains even after shell carbon
burning.
When the central temperature becomes 
$\log_{10} T_{{\rm C}} ({\rm K}) \sim 9.2$, the core
neon burning starts and an O/Si core forms (Fig.~\ref{fig_ev15}($c$)).
After the Ne burning, oxygen ignites at the center with the central temperature
of $\log_{10} T_{\rm C} ({\rm K}) \sim 9.3$.
The core oxygen burning produces a $\sim 1 M_\odot$ Si core 
(Fig.~\ref{fig_ev15}($d$)).
The main components of the Si core is $^{28}$Si, $^{32}$S, and $^{34}$S
and the electron fraction decreases by electron captures.
The Si core extends through the following shell oxygen burnings and the 
central temperature increases by the contraction.
When the central temperature is 
$\log_{10} T_{\rm C} ({\rm K}) \sim 9.55$, the Si burning
starts and an Fe core forms (Fig.~\ref{fig_ev15}($e$)).
The composition in the O layer also gradually changes through shell burnings.
After the Fe core formation, the Fe core extends with shell Si burning
and finally collapses to explode as a supernova.

Here we stress that one of the most important factors in the massive
star evolution is the carbon abundance after core helium burning, \x12c.
In general smaller carbon abundance leads a larger iron core at core collapse.
This is because when the carbon abundance is smaller, shell carbon 
burning is weaker after helium burning. Then the core is less
supported by the convective carbon shell burning and the time between
the central helium burning to the oxygen burning is shorter.
As a result, for smaller carbon abundance a larger core with higher entropy 
is formed because neutrino cooling is less effective\cite{WW86,NH88}.

 The carbon abundance also affects on nucleosynthesis substantially.
When the abundance is too low, carbon burning products such as Ne, Na, Mg, and Al
are underproduced compared with the solar abundance pattern, while they
are overproduced when the abundance is too large\cite{WW93}.
They also found that the abundances of S, Ar, and Ca are anti-correlates
with Ne, Na, Mg and Al. 

 Despite of these importance the carbon abundance is very uncertain
because it sensitively depends on the \cag rate and the treatment
of convection. Ref.~\citen{WW93}  
discussed that due to the nucleosynthetic constraint described above,
in their model
the \cag rate needs to be 1.7 $\pm$ 0.5 times the Caughlan and Fowler 
(1988) (CF88 hereafter) rate.\cite{CF88} 
In the NH88 model\cite{NH88} 
the rate of Ref.~\citen{CFHZ85} 
was taken which corresponds to $2.3-2.4$ times the CF88 rate.
In NH88, \x12c = (0.25, 0.22, 0.19) for the helium star
model for $M_{{\rm He}}$= (3.3, 6, 8) \msun, which roughly corresponds to the
ZAMS $M$=(13, 20, 25) \msun, respectively. 
In the UN model the \cag rate was chosen to be 1.3 times larger than 
the CF88 rate
so that the abundance pattern of EMP and VMP stars are 
reproduced.\cite{UN05,TUN07}

 Although there have been improvements in the estimation of the 
\cag rate\cite{Buchmann96,Kunz02}
uncertainty in the rate is still about factor of two. Therefore,
the nucleosynthetic method is still the best way to constrain the
carbon abundance. 

 We show in Table I the central carbon abundance after the core helium
burning for UN and YU models. 
In the YU model \cag rate is chosen to be 1.5 times as large as the CF88 rate
so that the carbon abundance
\x12c =0.20 for the $M$ = 25\msun ~model. As shown in the table,
mainly because of the difference in the \cag rate, the YU models
presented in this paper have systematically smaller carbon abundance than
the UN models. In Table II we show also final, $M_{\rm f}$, He core, 
$M_{\rm He}$, and
CO core masses, $M_{\rm CO}$. Here the latter two masses are defined where
the mass fraction of H and He are less than 0.001, respectively.

\begin{wraptable}{c}{\halftext}
\caption{Carbon mass fraction after core He burning for UN and YU models.
}
\label{table:1}
\begin{center}
\begin{tabular}{cccccccc} \hline \hline
UN model &   &  &&&&&\\ \hline
$M$($M_\odot$)  & 11.5  & 12 & 13 & 15 & 20 & 22 & 25 \\
\x12c   & 0.36  & 0.36& 0.33& 0.36& 0.34& 0.28&0.26\\
\hline\hline
YU model &   &  &&&&\\ \hline
$M$($M_\odot$)  & 11  & 12 & 13 & 15 & 18& 20 & 25 \\
\x12c   & 0.26  & 0.25& 0.31& 0.24&0.23 & 0.23&0.20\\
\hline
\end{tabular}
\end{center}
\end{wraptable}

\begin{wraptable}{c}{\halftext}
\caption{Final, core and remnant masses for UN, YU and NH88 models.
}
\label{table:2}
\begin{center}
\begin{tabular}{cccccccc} \hline \hline
UN model &   &  &&&&&\\ \hline
$M (M_\odot$)  & 11.5  & 12 & 13 & 15 & 20& 22 & 25 \\
$M_{\rm f} (M_\odot$)  & 11.2  & 11.6 & 12.7 & 13.9 & 17.5& 17.6 & 17.7 \\
$M_{\rm He}$ $(M_\odot$)  & 2.8  & 3.0 & 3.3 & 4.0 & 5.3& 6.7& 7.9\\
$M_{\rm CO}$ $(M_\odot$)  & 1.62  & 1.72 & 1.95 & 2.51 & 3.91& 5.42& 6.54\\
$M_{\rm rem}$ $(M_\odot$)  & 1.40  & 1.43 & 1.45 & 1.51 & 1.57 & 1.66& 1.70\\
$M_{\rm g}$ $(M_\odot$)  & 1.26  & 1.29 & 1.30 & 1.35 & 1.40 & 1.46& 1.50\\

\hline\hline
YU model &   &  &&&&&\\ \hline
$M (M_\odot$)  & 10 & 11 & 12 & 13 & 15& 18 & 20 \\
$M_{\rm f}$ $(M_\odot$)  & 9.5  & 10.5 & 11.4 & 12.1 & 13.6& 16.2 & 17.6 \\
$M_{\rm He}$ $(M_\odot$)  & 2.6  & 2.8 & 3.1 & 3.2 & 4.2& 5.4& 6.2\\
$M_{\rm CO}$ $(M_\odot$)  & 1.47  & 1.61 & 1.81 & 1.87 & 2.64& 3.58& 4.34\\
$M_{\rm rem}$ $(M_\odot$)  & 1.29  & 1.32 & 1.44 & 1.45 & 1.64& 1.77& 1.96\\
$M_{\rm g}$ $(M_\odot$)  & 1.18  & 1.19 & 1.29 & 1.30 & 1.45& 1.55& 1.69\\

\hline\hline
NH88 model &   &  &&&&&\\ \hline
$M (M_\odot$)  & $\sim$ 13 & $\sim$ 15 & $\sim$ 20 & $\sim$ 25 & &  &  \\
$M_{\rm He}$ $(M_\odot$)  & 3.3  & 4 & 6 & 8 & & & \\
$M_{\rm rem}$ $(M_\odot$)  & 1.27  & 1.33 & 1.61 & 1.77 & & & \\
$M_{\rm g}$ $(M_\odot$)  & 1.15  & 1.20 & 1.43 & 1.55 & & & \\

\hline
\end{tabular}
\end{center}
\end{wraptable}

%
%

\subsection{Evolution of 10-13 $M_\odot$ stars}

\begin{figure}[tbp]
\centerline{
\includegraphics[width=8cm,angle=270]{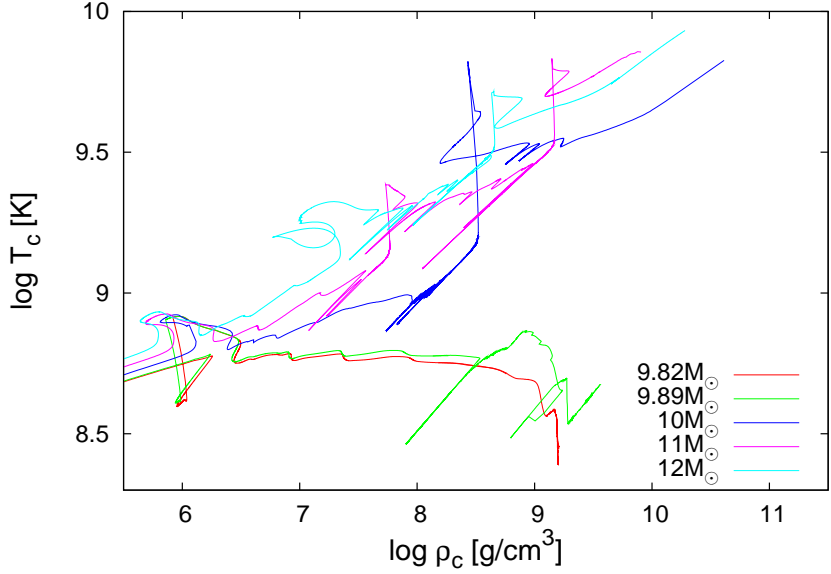}}
\caption{Evolution tracks of $M = 9.82, 9.89, 10, 11$, and $12 M_\odot$ 
stars in terms of central temperature and density.
}
\label{fig_KT1}
\end{figure}

Here we describe the evolution of 10-13 $M_\odot$ stars calculated by
the YU code since these stars correspond roughly to the lightest Fe core
collapse model, and the evolution after oxygen burning is quite different
from more massive $M > 13 M_\odot$ stars.
Table \ref{tbl_KT1} shows the differences in the way how the stars commence
nuclear burning of Ne, O, and Si.
In the table
a letter `C' represents that the ignition occurs at the center of the core
and `off-C' shows that
the ignition occurs off-centrally.
In the line of the composition, letters represent main nuclei at the center.
The differences also appear in Fig.~\ref{fig_KT1}, in which evolutional
tracks are shown in terms of the central temperature and density.
In this mass range, a star forms a CO core around the Chandrasekhar
limiting mass. The degeneracy of the core is the important property
and affects its evolution.

We will present more detailed description in Takahashi, Yoshida \& Umeda 
(2012, in preparation, see also Ref. \citen{TUY12}),
in which stellar evolutions are calculated in very fine grids. So far,
we have found that $M = 9.89 M_\odot$ model ends up the Fe core formation,
though $M = 9.82 M_\odot$ model does not form an Fe core.
As described below the off-center Si-burning is more
and more violent for less massive stars. In this mass range only 0.01 $M_\odot$
difference in the zero-age main sequence mass would result in quite different evolution.
However, the global evolutionary properties and final density structure
of an Fe core is similar to the 10 $M_\odot$ model presented here as far as
an Fe core is formed.

{\underline{\it 10 $M_\odot$ star}:}
For a 10 $M_\odot$ star, the evolution from the zero age main sequence 
to the C burning phase goes 
roughly in the same way as more massive stars which end up with the Fe core
formation. However, after an O-Ne core formation through the C
burning, different evolutionary aspects come about.

\begin{figure}[tbp]
\centerline{
\includegraphics[width=8cm,angle=270]{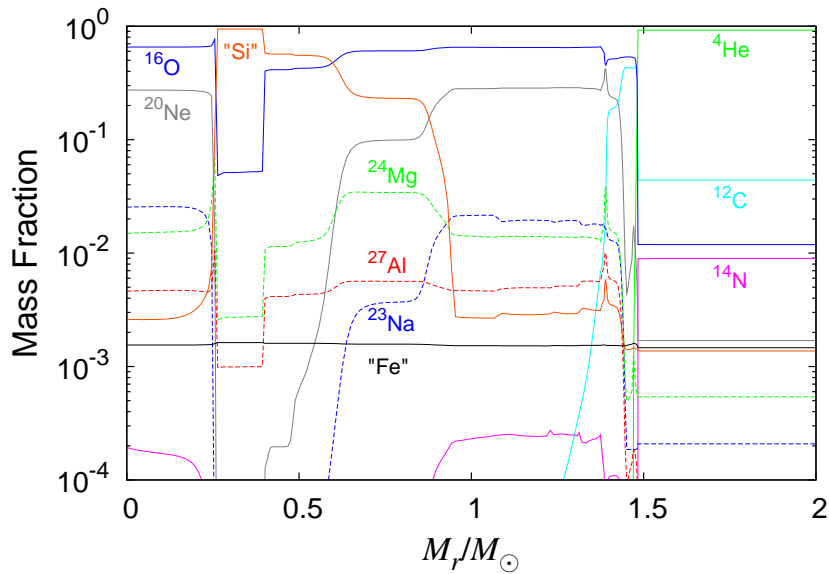}}
\caption{Mass fraction distribution for a 10$M_\odot$ star
during a shell Ne+O burning, when $\log_{10}\rho_{\rm C} = 7.9358$.
}
\label{fig_KT2}
\end{figure}

Because the mass of the C-O core of 1.47$M_\odot$ is only slightly larger
than the critical mass for the Ne
ignition, the O-Ne core first contracts without nuclear burning.
Eventually, the importance of the pressure of degenerate electron increases.
Such semi-degenerate cores often show the inverse temperature distribution;
temperature does not take its highest value at the center.
This is due to the property of degenerate electrons.
Since neutrino cooling is more
effective at higher density as well as at higher temperature, the
O-Ne core loses its heat by
the cooling. As heat escapes, the core contracts and compressional heating
supply energy into the
gas of ions, electrons, and radiation. If the core
is supported by pressure of ions, 
gas temperature increase as a result of compressional
heating. On the other hand, if electron pressure supports the core,
gas temperature is hard to increase because internal energy of degenerate electrons
does not depend on temperature but on density.
What the high density O-Ne core supports mainly is the pressure of degenerate electrons, so
the denser the region is, the slower the temperature increase is.
Owing to this temperature inversion, a shell Ne burning commences at 
an off-center region having
higher temperature than the center. Since Ne and O have close
ignition temperatures, once the shell Ne burning commences, heat
generation increases temperature and the shell O-ignition succeeds.
Then a degenerate O-Ne core is surrounded by a hot 
Si-shell (Fig.\ref{fig_KT2}).
These evolutionary properties are fully consistent with previous
works (e.g., Refs. \citen{NH88,Poelarends08}).

\begin{figure}[tbp]
\centerline{
\includegraphics[width=8cm,angle=270]{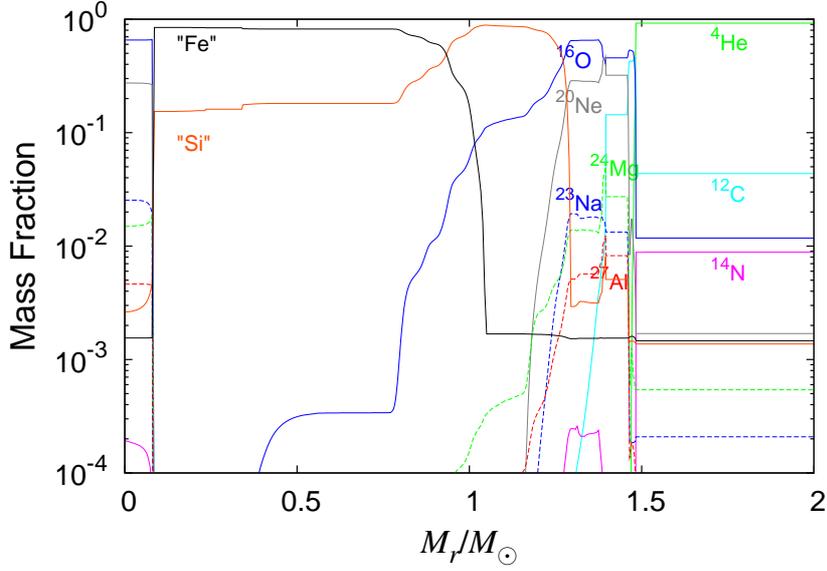}}
\caption{Mass fraction distribution for a 10$M_\odot$ star
during a shell Ne+O+Si burning, when $\log_{10}\rho_{\rm C} = 8.0958$.
}
\label{fig_KT3}
\end{figure}

Some of the nuclei of the Si cluster produced by the O burning can capture
electrons in such a high density O-Ne core. Since the core is mainly
supported by the pressure of degenerate electrons, the decrease of the number
of electrons causes the core contraction. Heat production by the contraction
increases the temperature. When temperature at the base of the Si shell
becomes high enough to ignite Si, the shell Si burning commences and transforms
the Si shell into the Fe shell (Fig. \ref{fig_KT3}).
The energy production rate of the burning from Ne to Fe is
very high. So at every ignition, the core repeats the expansional cooling and
compressional heating almost adiabatically,{\footnote{This is seen in spikes around $\log_{10}\rho_{\rm C} \sim 8$ in Fig.~\ref{fig_KT1}.}}
and the burning front approaches the center little by little.
After the burning front reaches the center of the core, an Fe core forms.
Then the star collapse.
Some vertical ascents in Fig.~\ref{fig_KT1} show moments when a base of
a off-center burning reaches the center of the core.
For a 10$M_\odot$ star, the central temperature suddenly increase at
$\log_{10} \rho_{\rm C}=8.4319$ when the shell Ne+O+Si burning reaches 
the center.


{\underline{\it 11 $M_\odot$ star}:}
For a 11 $M_\odot$ star with a larger C-O core of 1.61$M_\odot$ than a 10$M_\odot$
star, electron degeneracy of the core is less. This causes the difference in
evolution after Ne and O ignite off-centrally.

\begin{figure}[tbp]
\centerline{
\includegraphics[width=8cm,angle=270]{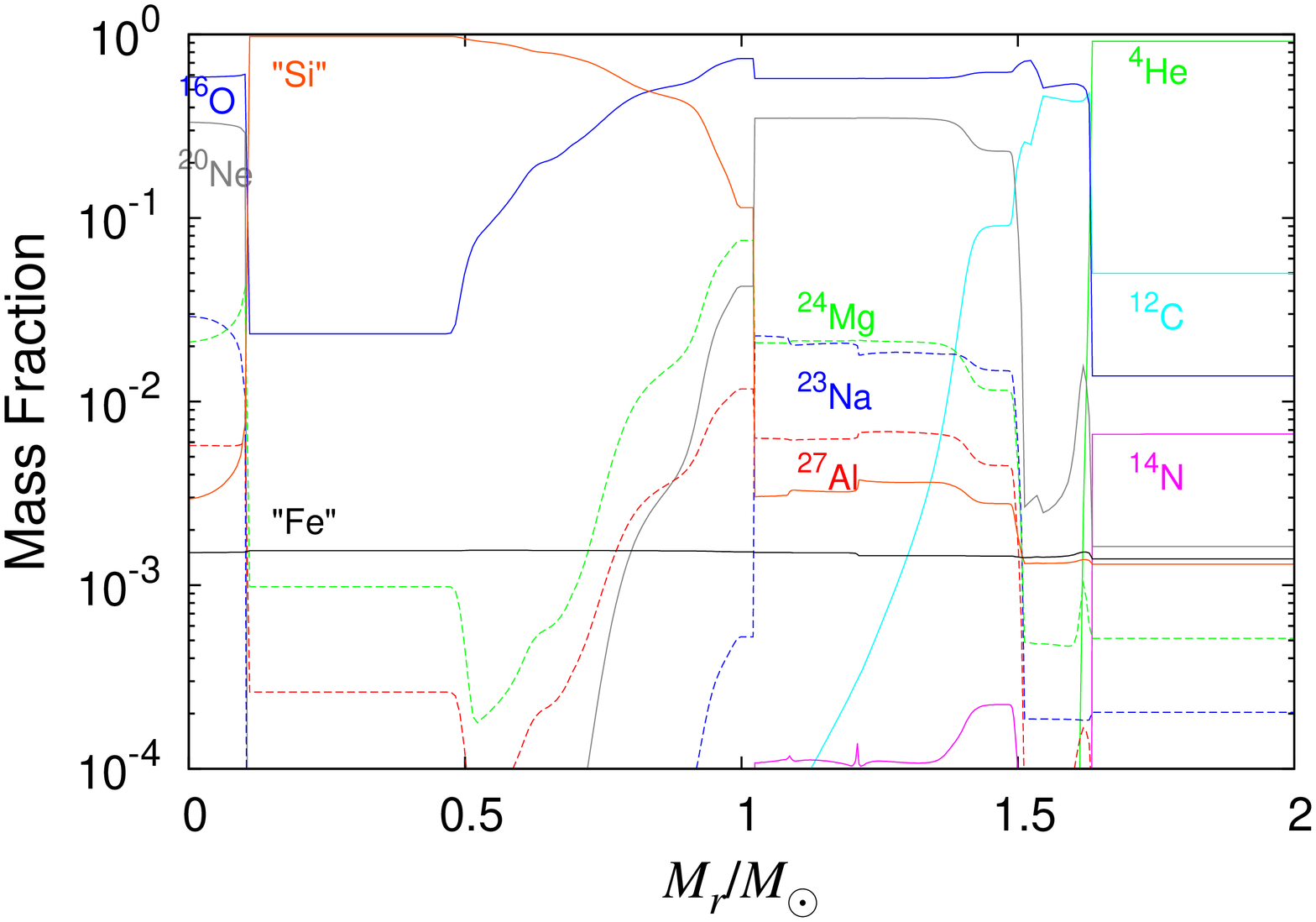}}
\caption{Mass fraction distribution for a 11$M_\odot$ star
during a shell Ne+O burning, when $\log_{10}\rho_{\rm C} = 7.6181$.
}
\label{fig_KT4}
\end{figure}

Because of the less degeneracy at the center, a difference
between the maximum and the central temperature 
is less, so a shell O+Ne ignition commences at
deeper region than the 10$M_\odot$ star (Fig. \ref{fig_KT4}).
The temperature at the base of the shell increases by compressional 
heating as the same way as the 10$M_\odot$ star during the O+Ne burning front 
propagation.
However, the burning front arrives at the center before the base temperature
reaches the Si ignition and, thus, a Si core forms.
This is the main difference between 10 and 11$M_\odot$ stars.

\begin{figure}[tbp]
\centerline{
\includegraphics[width=8cm,angle=270]{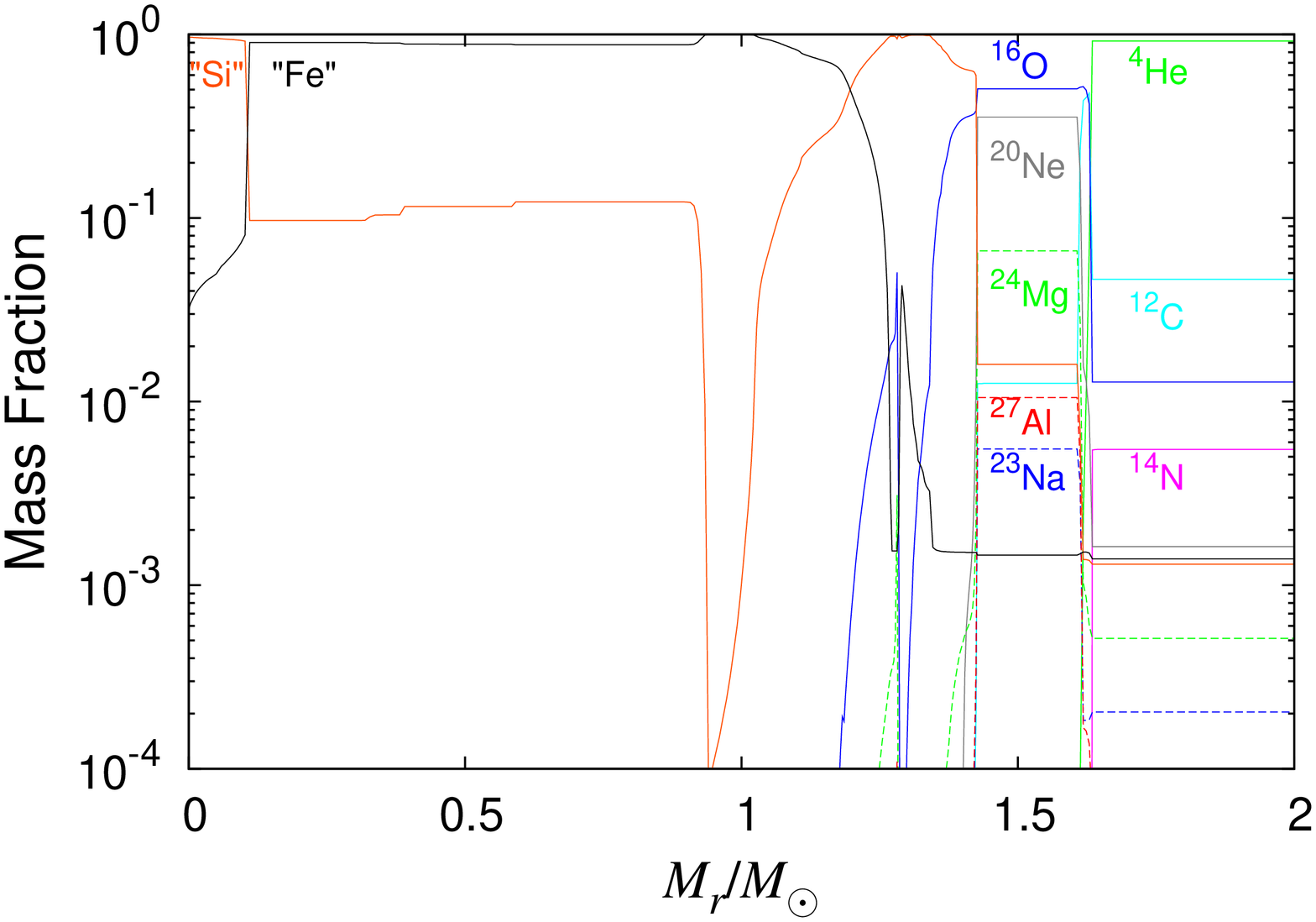}}
\caption{Mass fraction distribution for a 11$M_\odot$ star
during a shell Si burning, when $\log_{10}\rho_{\rm C} = 8.0481$.
}
\label{fig_KT5}
\end{figure}

Later, as the Si core contracts,
outer O+Ne shell burning repeats moving its base outward,
while neutron rich nuclei such as $^{36}$S and $^{50}$Ti are produced in inner region.
Degeneracy of electrons also increases in the contracting core and 
the above-mentioned temperature inversion appears.
When the central density reaches the value of $\log_{10}\rho_{\rm C} = 8.9867$,
the first off-center Si ignition occurs at $M_r = 0.5933M_\odot$.
This ignition causes the core to expand adiabatically.
After the burning terminates, the core contracts adiabatically again.
When the central density becomes $\log_{10}\rho_{\rm C} = 9.0253$, 
the second off-center Si ignition occurs at $M_r = 0.1103M_\odot$
(Fig. \ref{fig_KT5}).
After the second off-center Si ignition,
the burning front reaches the center.
The core transforms into an Fe core, then collapses.


{\underline{\it 12 $M_\odot$ star}:}
A 12$M_\odot$ star evolves similar to a 11$M_\odot$ star
but the properties of Ne and O ignitions are different.
Because it has a larger C-O core of 1.85$M_\odot$,
temperature takes its highest value at the center of the O-Ne core.
Therefore, Ne and O ignite at center.
Note that Table \ref{tbl_KT1} shows that when O ignites at the center,
central
composition is O-Si, and not O-Ne. This is just because Ne has burned
ahead of O, producing a O-Si core.

\begin{figure}[tbp]
\centerline{
\includegraphics[width=8cm,angle=270]{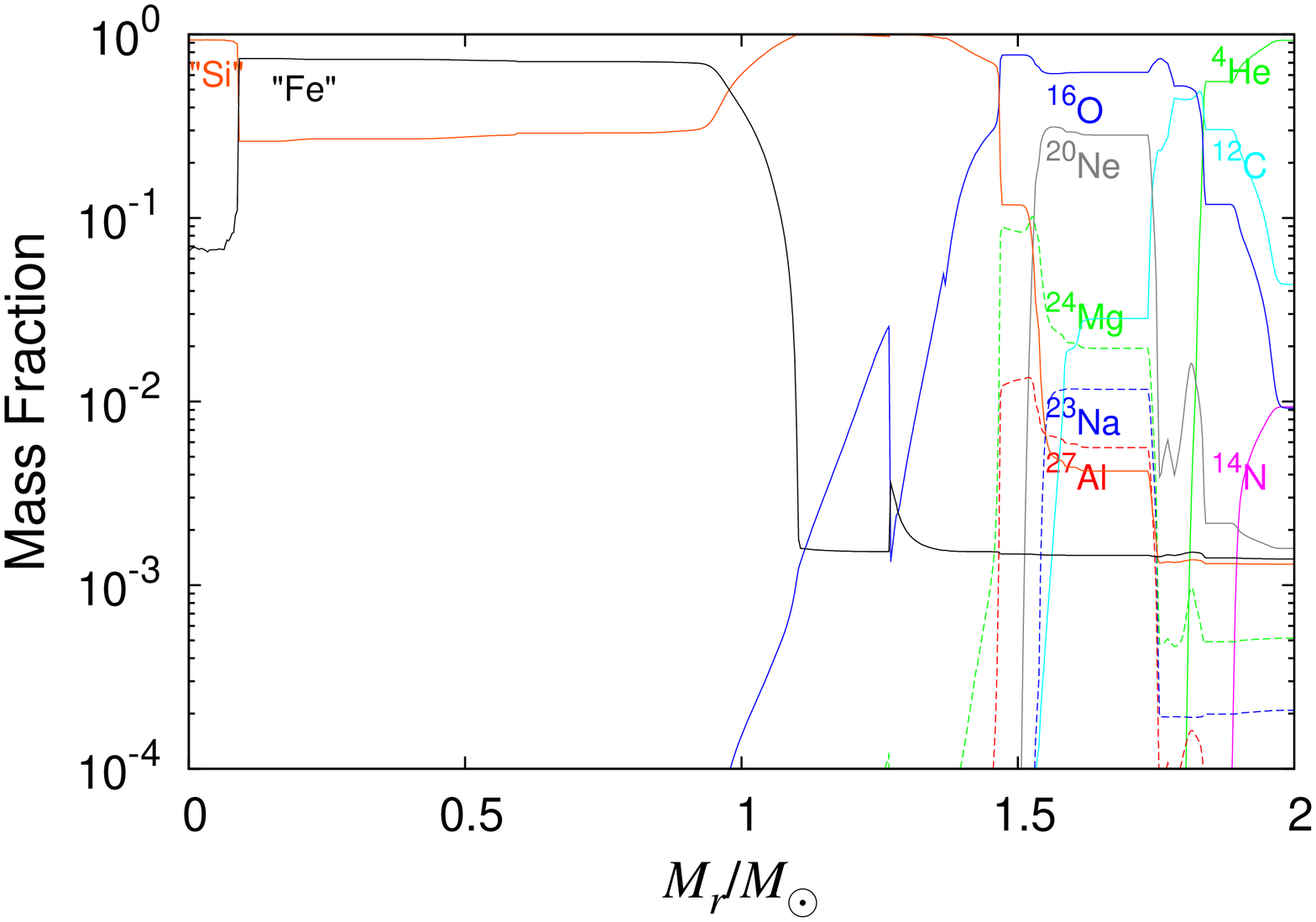}}
\caption{Mass fraction distribution for a 12$M_\odot$ star
during a shell Si burning, when $\log_{10}\rho_{\rm C} = 7.9694$.
}
\label{fig_KT6}
\end{figure}

After Si core formation, the evolution proceeds as the same way as
a 11$M_\odot$ star.
Neutron rich nuclei are produced around the center after
several shell Ne+O burning at outer regions of the contracting core.
Then off-center Si burning occurs (Fig. \ref{fig_KT6}),
and after the front reaches the center an Fe core forms. 

\begin{wraptable}{c}{\halftext}
\label{table:3}
\caption{Ignition properties of $10 - 13 M_\odot$ stars.
}
\label{tbl_KT1}
\begin{center}
\begin{tabular}{ccccc} \hline \hline
C ignition & 10$M_\odot$ & 11$M_\odot$ & 12$M_\odot$ & 13$M_\odot$ \\ \hline
at the center or not  & C & C & C & C \\
central composition   & C-O & C-O & C-O & C-O \\
 \hline \hline
Ne ignition & 10$M_\odot$ & 11$M_\odot$ & 12$M_\odot$ & 13$M_\odot$ \\ \hline
at the center or not  & off-C & off-C & C & C \\
central composition   & O-Ne & O-Ne & O-Ne & O-Ne \\
 \hline \hline
O ignition & 10$M_\odot$ & 11$M_\odot$ & 12$M_\odot$ & 13$M_\odot$ \\ \hline
at the center or not  & off-C & off-C & C & C \\
central composition   & O-Ne & O-Ne & O-Si & O-Si \\
 \hline \hline
Si ignition & 10$M_\odot$ & 11$M_\odot$ & 12$M_\odot$ & 13$M_\odot$ \\ \hline
at the center or not  & off-C & off-C & off-C & off-C \\
central composition   & O-Ne & Fe-Si & Fe-Si & Fe-Si \\
\hline
\end{tabular}
\end{center}
\end{wraptable}


\begin{figure}[p]
\centerline{
\includegraphics[width=7cm,angle=270]{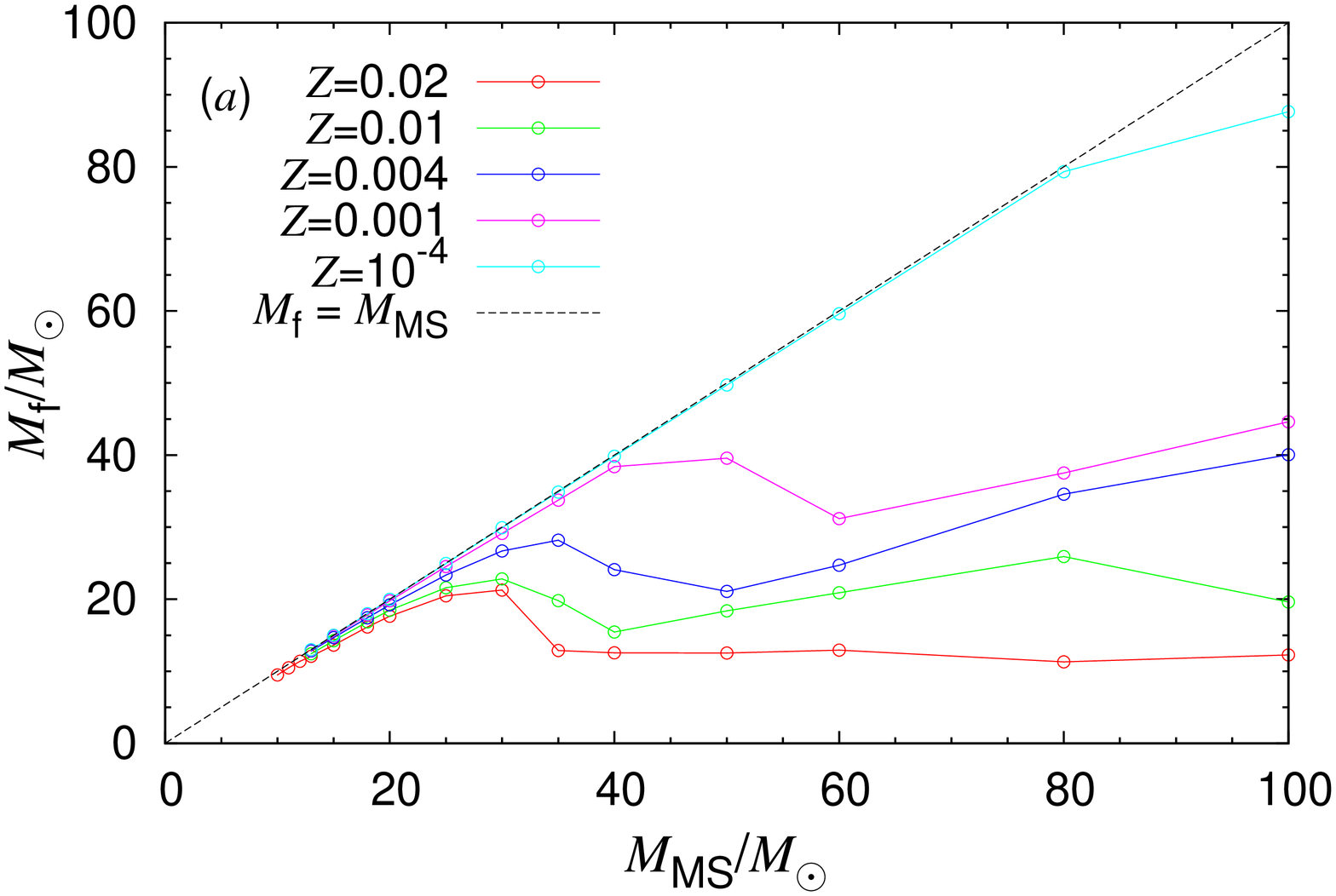}}
\centerline{
\includegraphics[width=7cm,angle=270]{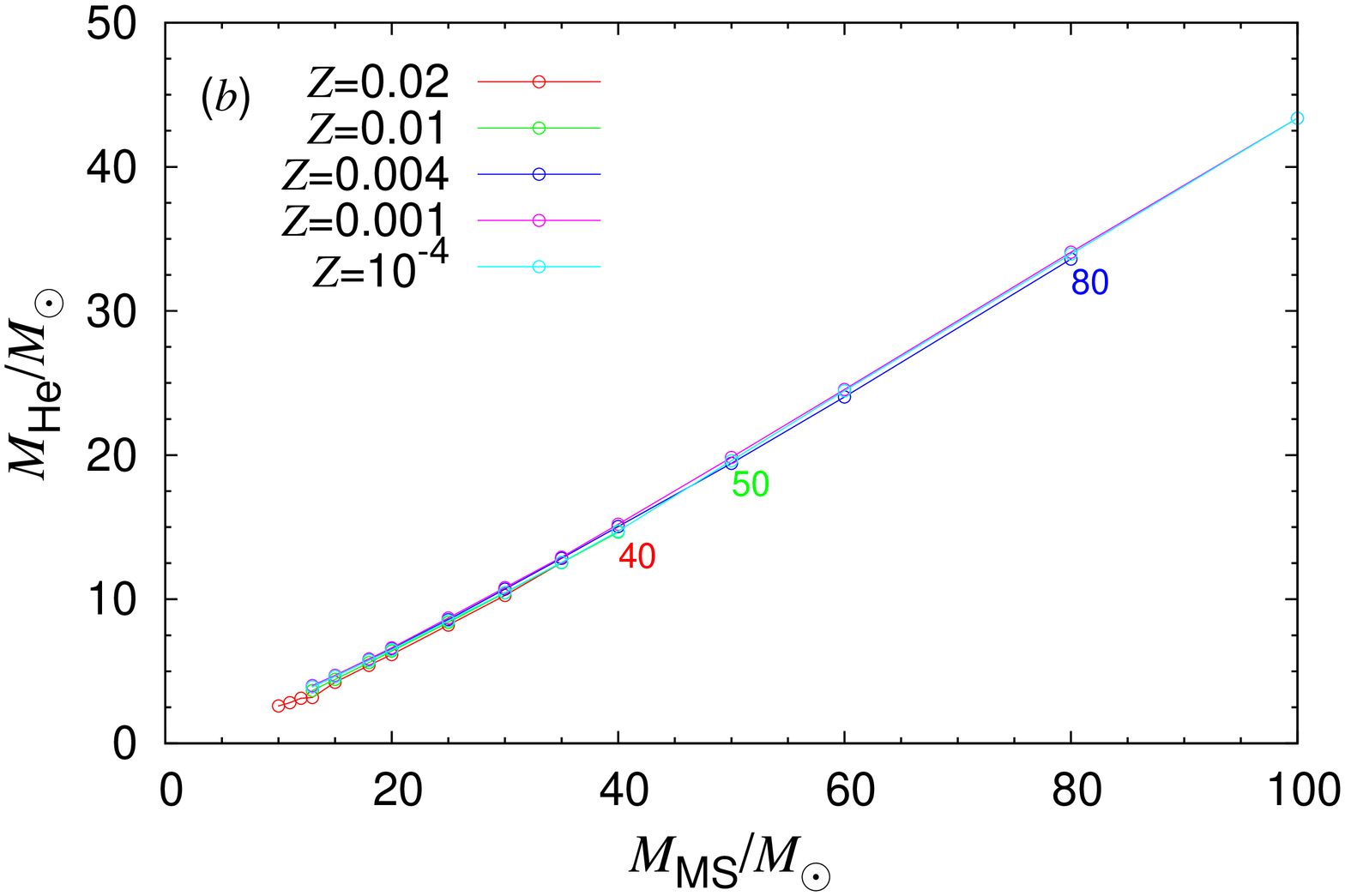}}
\centerline{
\includegraphics[width=7cm,angle=270]{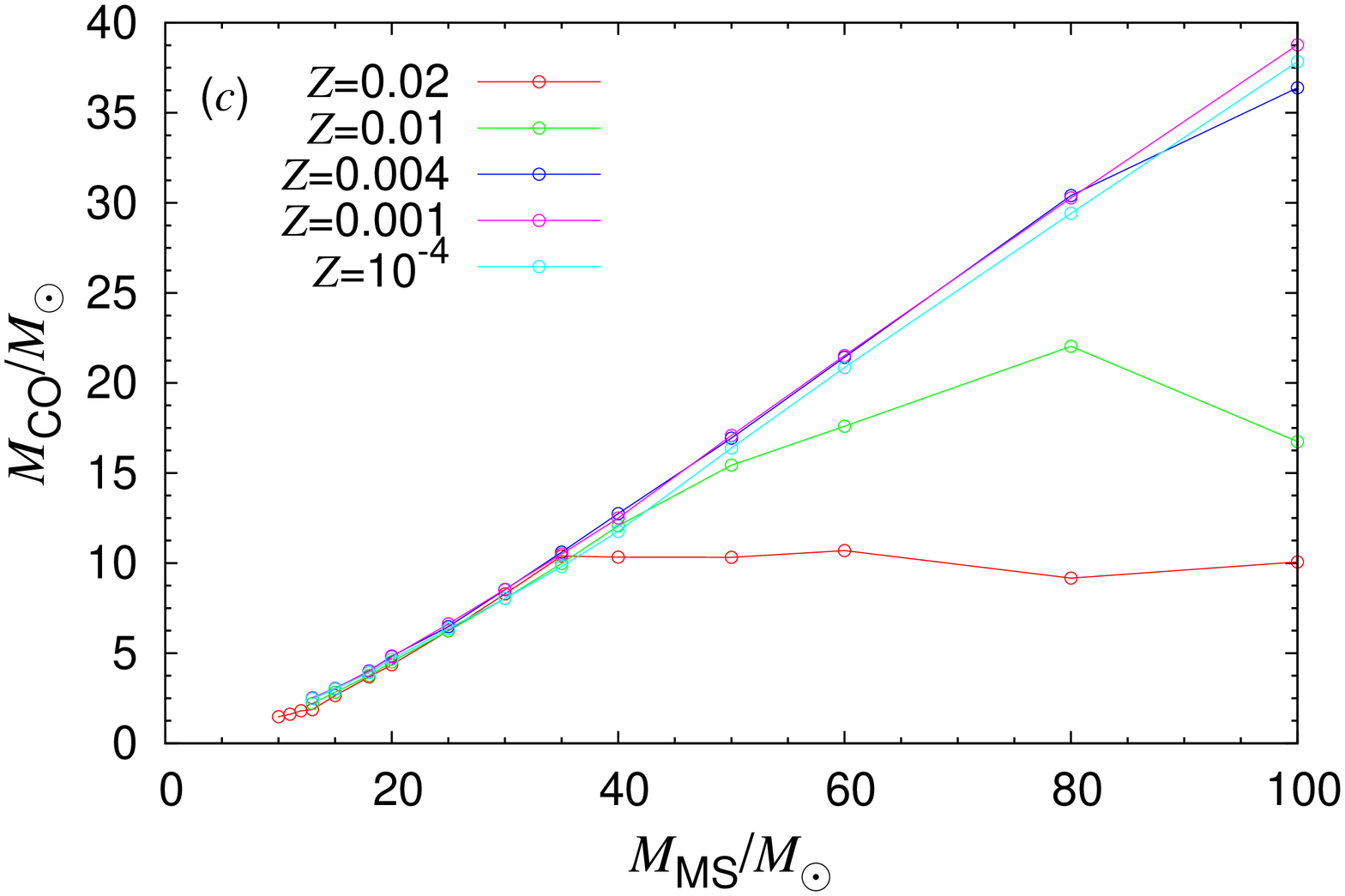}}
\caption{The final mass (panel ($a$)), the He core mass (panel ($b$)), 
and the CO core mass (panel($c$)) with the relation to the main-sequence mass.
The adopted metallicities are $Z = 10^{-4}, 0.001, 0.004, 0.01$, and 0.02.
In panel ($b$), stars more massive than the attached number of the 
main-sequence mass ($M_\odot$) evolve to WO stars.
The surface He mass fraction of the stars is about 0.2.
}
\label{fig_Mf}
\end{figure}

\subsection{Final mass and metallicity}

 In Fig.~\ref{fig_Mf} we show the metallicity dependence of the
final and core masses of the YU models\cite{YU11}.
We see clear metallicity dependence in the final mass among the stars
with $M \gsim 20 M_\odot$.
In $Z = 0.02$, $M = 30 M_\odot$ star has the maximum final mass:
$M_{\rm f} = 21.3 M_\odot$.
$M \gsim 40 M_\odot$ stars indicate a roughly constant final
mass of $\sim 10 M_\odot$.
These stars lose all of H and He layers and become Wolf-Rayet stars.
Effect of mass loss in $Z = 0.01$ stars is less than that in
$Z = 0.02$ stars.
Increasing the main-sequence mass, the final mass also increases in
$M \le 30 M_\odot$ and $40 \le M \le 80 M_\odot$
but it decreases in $30 \le M \le 40 M_\odot$ and 
$M \ge 80 M_\odot$.
Metal poorer stars have similar $M$ dependence of the
final mass.
In this figure the zero-age main-sequence mass where the mass loss becomes 
effective and the final mass for a given main-sequence mass become larger 
for metal poorer stars.
We do not see clear mass loss effect among $M \le 80 M_\odot$ stars with
$Z = 10^{-4}$.

 The metallicity dependence of the He core is small (Fig.~\ref{fig_Mf}($b$)).
On the other hand, most of the He layer has lost during the evolution
in $M \ge 40, 50$, and 100 $M_\odot$ stars for 
$Z = 0.02, 0.01,$ and 0.004.
These stars lose the He/C/O envelope and their surface He abundance
decreases during the He burning.
They become C- and O-enriched Wolf-Rayet stars (WO stars) with the
surface He mass fraction of $\sim 0.2$.
The mass of the He-rich shell is $1.4 - 5.5 M_\odot$ in $Z = 10^{-4}$ stars,
in which mass loss effect is small.
Since the fraction of the He layer is small, mass loss brings about the 
removal of He layer rather than the reduction of the He core mass.

 The metallicity dependence of the CO core mass is seen in $Z > 0.004$
and $M \gsim 40 - 50 M_\odot$ in Fig.~\ref{fig_Mf}($c$).
The CO core mass is roughly constant with $\sim 10 M_\odot$ in 
$Z = 0.02$ and $M \ge 40 M_\odot$ stars.
In $Z = 0.01$, the maximum mass of the CO core is $22 M_\odot$ of 
$M = 80 M_\odot$ star.
On the other hand, the CO core mass can be more massive in metal-poorer
stars. The CO core mass is $35 - 40 M_\odot$ in
$Z \le 0.004$ and $M = 100 M_\odot$.

\section{Nucleosynthesis}

\begin{figure}[t]
\centerline{
\includegraphics[width=8cm,angle=270]{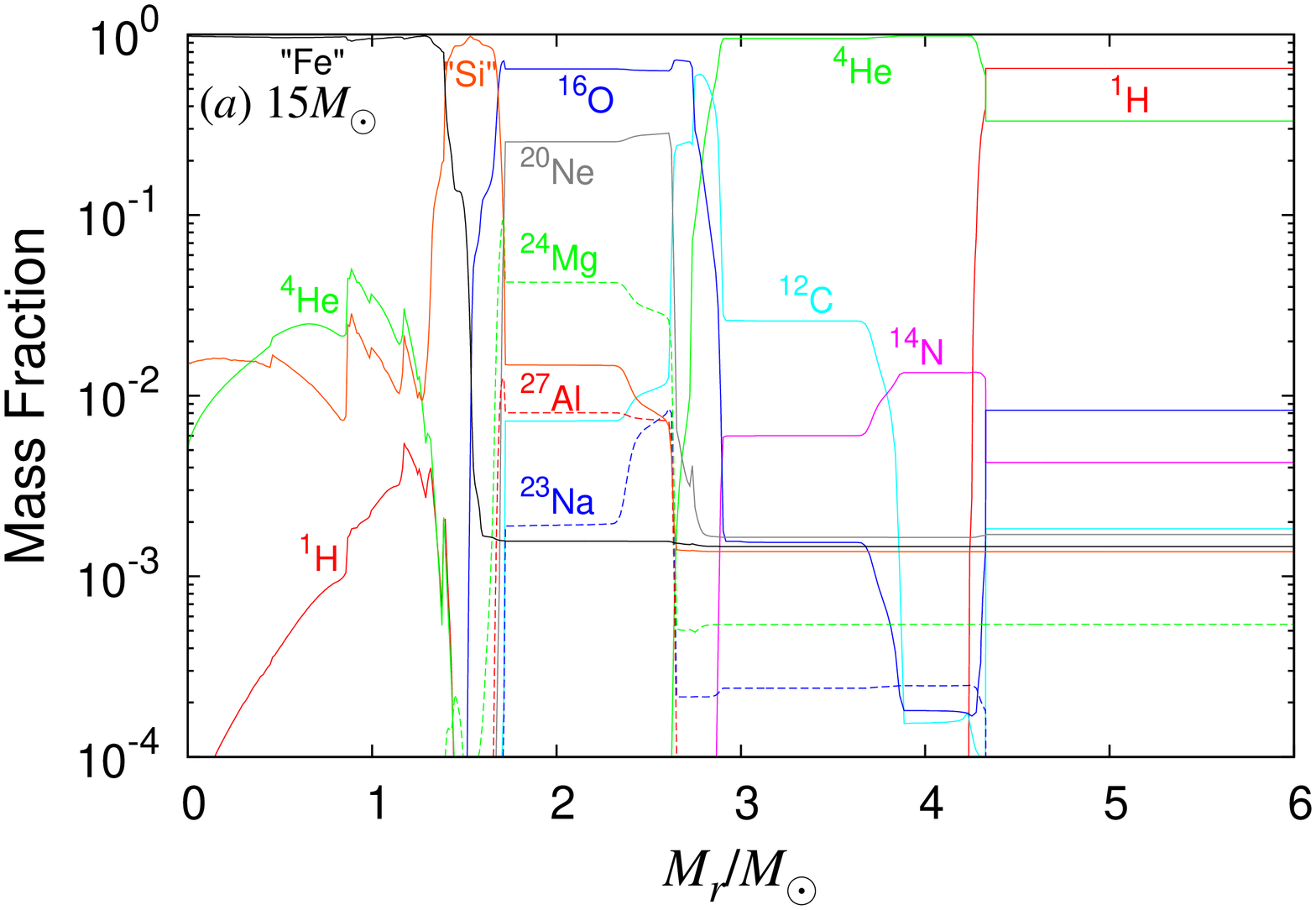}
}
\centerline{
\includegraphics[width=8cm,angle=270]{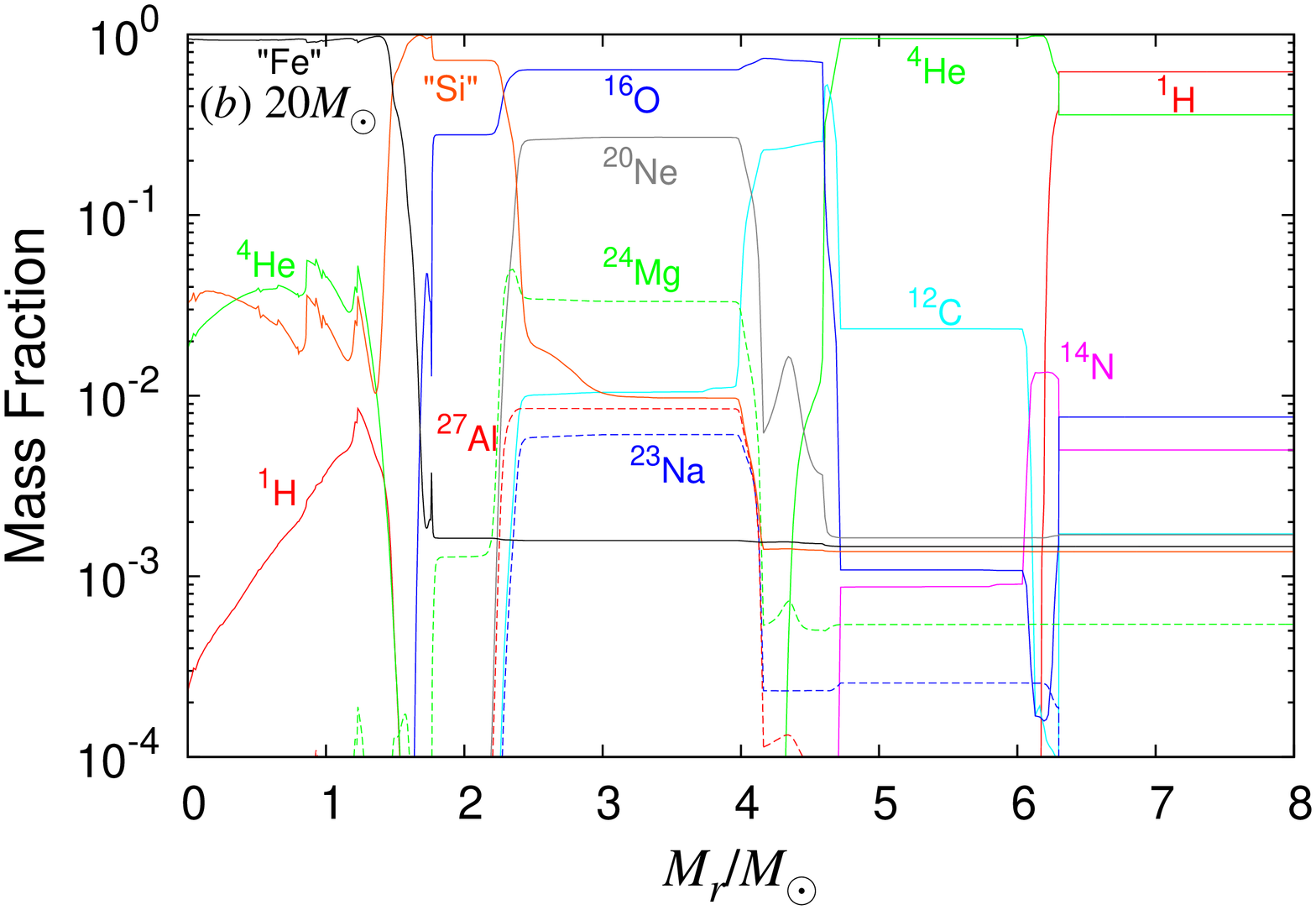}
}
\caption{Mass fraction distribution of the supernova progenitors evolved
from 15 ($a$) and 20 $M_\odot$ ($b$) stars with the metallicity of $Z = 0.02$.
The corresponding final masses are 13.6 and 17.6 $M_\odot$.
The mass fraction distribution in the H-rich envelope outside the range
of each figure is the same as that at the outer edge of the figure.
}
\label{fig_progmf}
\end{figure}

 Nucleosynthesis in these massive stars occurs mainly in two
stages. One is before core-collapse and the other is during supernova
explosion. It is well known that inside these massive stars nuclear fusion
takes place up to Fe synthesis. Before gravitational core collapse,
Fe core is produced at center and lighter elements form 
onion like structures from inside to the surface (Fig.~\ref{fig_progmf}).

 Nucleosynthesis during the supernova explosion is usually called
{\it explosive nucleosynthesis}. During the explosion, shock wave propagates
out of Fe-core to the stellar surface. Behind the shock wave
matter heats up and nuclear burning takes place. 
Explosive nucleosynthesis is important for the synthesis of
Si and heavier elements (e.g., Ref.~\citen{UN02}).

 In our previous works 
(e.g., Refs.~\citen{UN02,UN05,UN08,YUN08})
we use a simple 1D model for supernova explosion to calculate explosive
nucleosynthesis. We inject thermal or kinetic energy below the mass cut
or just above the Fe core to initiate supernova shock.
We may call this procedure as 'instant energy injection'.

 We use a 1D PPM code for the hydrodynamical calculations and solve 
small alpha-network 
together to calculate nuclear energy generation. Then we calculate
detailed nucleosynthesis post-processingly by solving a large
nuclear reaction network.

 Explosive nucleosynthesis should depend on how the star explodes,
but we do not know yet how gravitational collapse leads
to the explosion. Therefore, there are still several proposals for the
successful supernova explosions. Fortunately the 
properties of supernova shock outside the Fe core is almost independent of the
explosion mechanism, because in most models
the supernova shock gains energy inside the Fe core.
Then, as far as spherical symmetry is assumed, explosion energy $E$ is the
only parameter which determines the properties of the supernova shock.
For example, kinetic to thermal energy ratio
does not much affect the propagation of the shock outside the Fe core,
because it quickly converges into the same solution.
In this case 'instant energy injection' method is sufficient to
calculate nucleosynthesis. For most cases, nucleosynthesis of 
Fe peak elements including Zn and lighter elements can be safely calculated
with this method. In Section 7,
We will discuss the limitations of this simple procedure.

 Behind the shock front, temperature distribution is
almost homogeneous, and the radiation energy is strongly dominated.
Then the radiation temperature $T$ can be related to the explosion energy, $E$, 
by $E= (4 \pi/3) a r^3 T^4$, where $a$ is the radiation constant and
$r$ is the shock radius. 
Thus the maximum temperature after shock passage, $T_{\rm s}$, 
for the mass elements 
located at radius $R$ in the progenitor is approximately given 
as \cite{TNH96} 

\begin{equation}
  T_{\rm s} = (3E/ 4 \pi a R^3)^{1/4}.
  \label{eq:1}
\end{equation}

 For a given progenitor model and explosion energy $E$, the propagation
of shock and thus the time evolution of density distribution is also 
determined uniquely. Behind the shock front, pressure and entropy
are also radiation dominated and can be written as 
$P_\gamma  = a T^4/3$ and $S_\gamma = (4/3) a T^3/\rho$.
Since the shocked region cools roughly adiabatically at first, the radiation 
entropy is often convenient to specify the explosion.

\begin{figure}[t]
\centerline{
\includegraphics[width=8cm,angle=270]{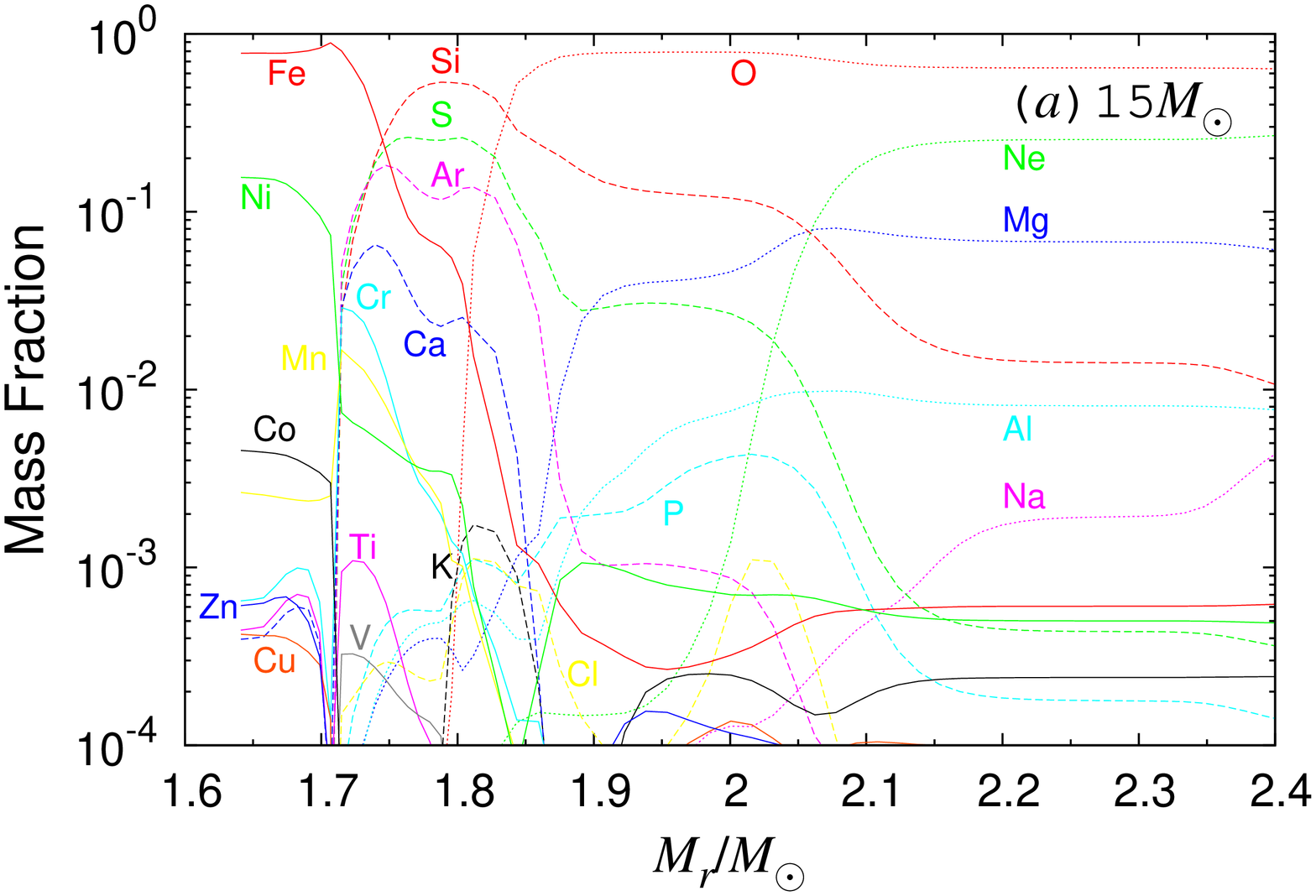}
}
\centerline{
\includegraphics[width=8cm,angle=270]{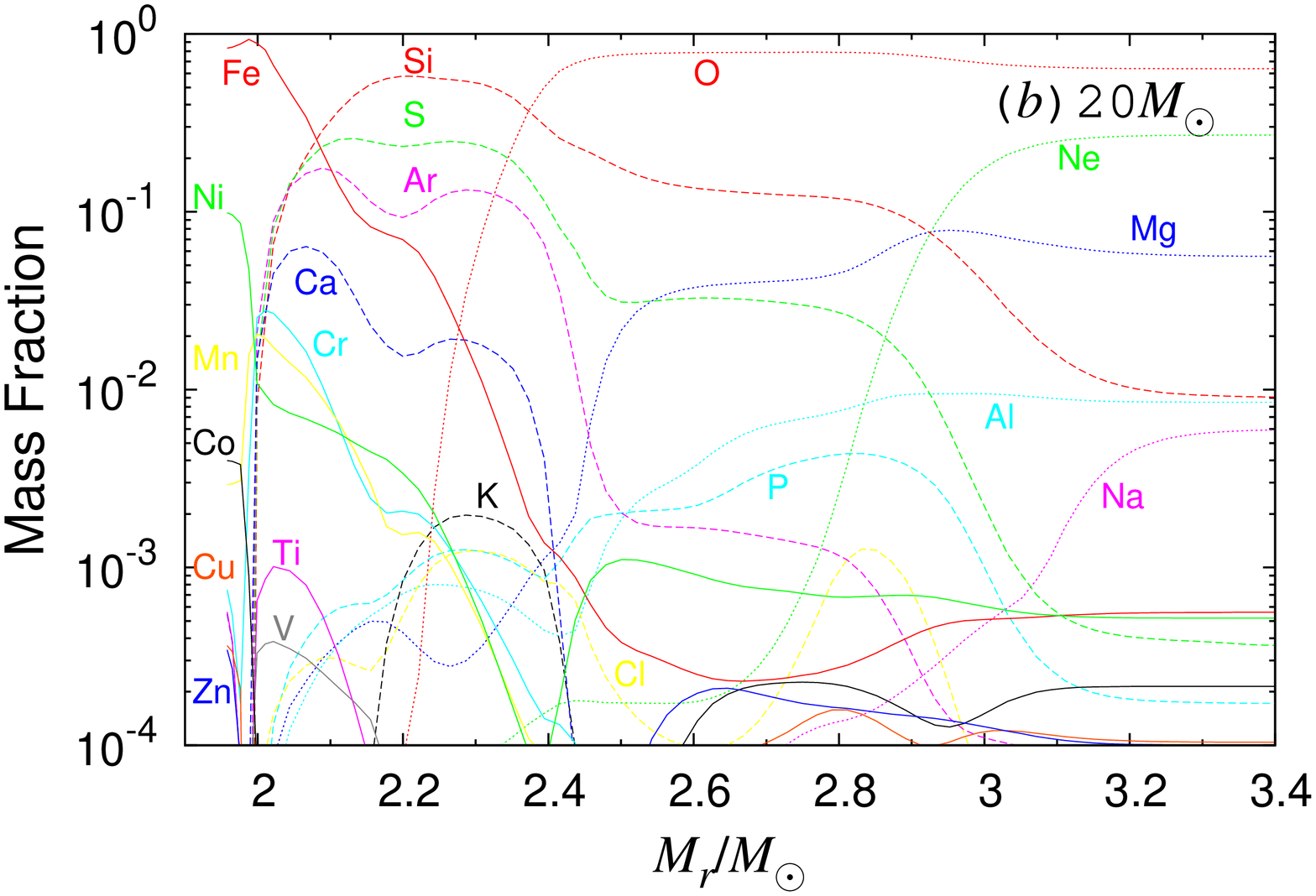}
}
\caption{Mass fraction distribution of the innermost region of the
supernova ejecta evolved from from 15 ($a$) and 20 $M_\odot$ ($b$) stars 
with the metallicity of $Z = 0.02$.
The explosion energy is set to be $1 \times 10^{51}$ erg.
}
\label{fig_expmf}
\end{figure}

  In the inner most region of supernova ejecta where $T_{\rm s}$ is 
higher than
$5 \times 10^9$ (K), Si burns completely. In such a region, at first Si mostly 
decomposes 
to alpha-particles, and this is endothermic reaction.  
Then as the temperature decreases alpha particle starts to recombine
and alpha-rich freezeout nucleosynthesis takes place. This phase is exothermic.
The energy changes due to these processes are typically about ten percent,
and thus above equation (\ref{eq:1})
still provides a useful approximation.

 In this 'complete Si burning' region $^{56}$Ni is dominantly produced 
but also Co, Ni, and Zn are mostly produced here. 
The final abundance depends on entropy because the mass fraction
of alpha-particles during the alpha-rich freezeout phase 
increases with entropy\cite{TNH96}. 

 For example, one of the authors has shown that high entropy
during the hypernova explosion explains well the larger
ratio of [Zn/Fe] and [Co/Fe] towards lower metallicity, [Fe/H],
in extremely metal poor (EMP) stars
\cite{UN02,UN05,TUN07,IUT09}.
They also discussed that simple 1D model cannot explain the abundance of
EMP stars by the hypernova models, because Fe or $^{56}$Ni is over-ejected 
to satisfy the large 
[Zn/Fe] ratio. This problem and the solution is described in the 
subsection 4.3.

 When $T_{\rm s}$ is between $4 - 5 \times 10^9$ (K), Si burns incompletely.
This region is called incomplete Si burning region. Main products here
is Cr, Mn, and $^{56}$Ni. 
For $T_{\rm s} \simeq$ 
$3 - 4 \times 10^9$ (K), oxygen burns partially
and Si, S, Ar, Ca, Ti, V are main products. In these regions also nuclear 
burning produces some amount of energy. 
In summary, previous works have shown that explosive nucleosynthesis
up to Zn can be described well by instant energy injection models,
though hypernova models over-produce Fe in simple spherically symmetric 
models.

\begin{figure}[t]
\centerline{
\includegraphics[width=8cm,angle=270]{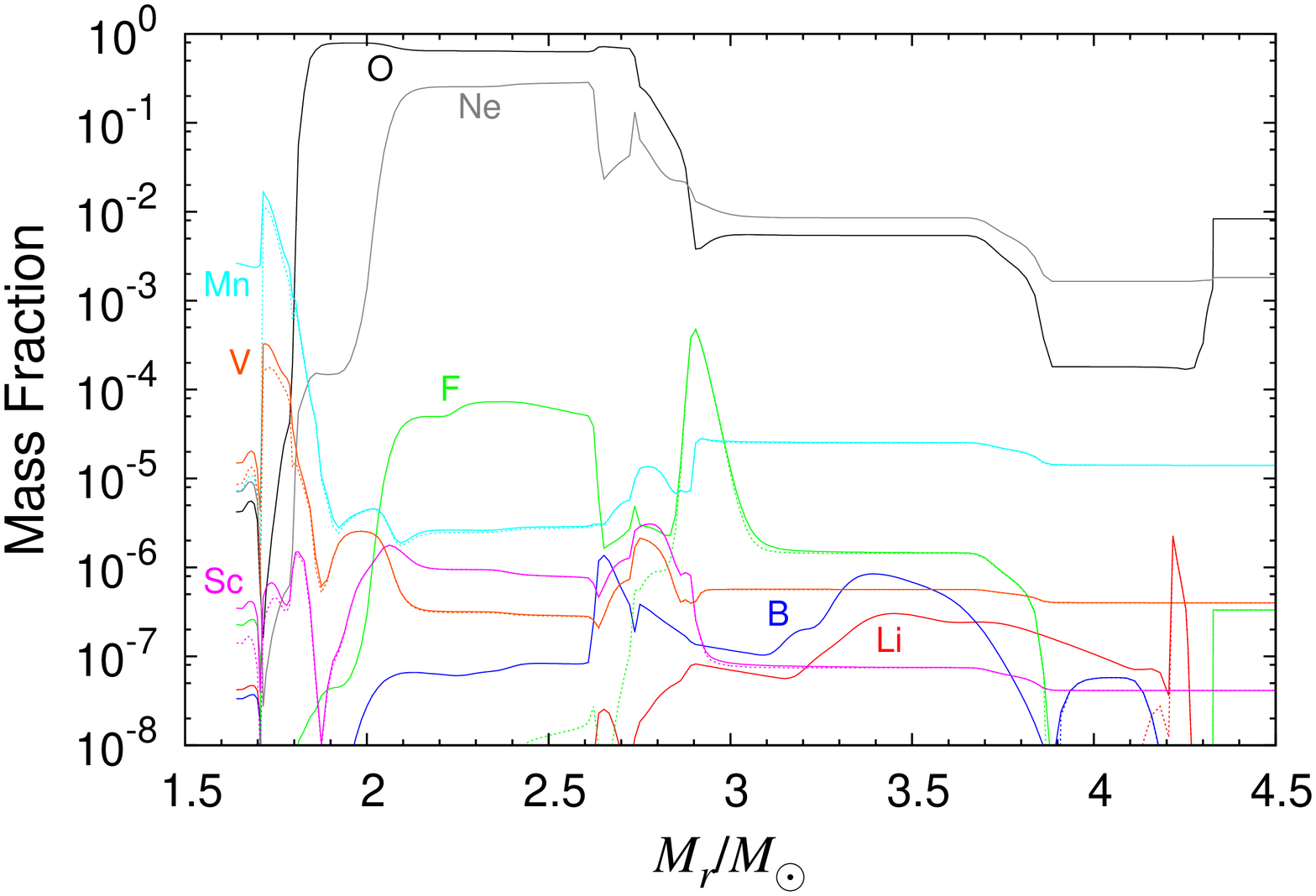}
}
\caption{Mass fraction distribution of Li, B, F, Sc, V, and Mn produced
through the $\nu$-process in the supernova ejecta evolved
from a 15 $M_\odot$ star with the metallicity of $Z = 0.02$.
Solid and dotted lines indicate the mass fractions of the elements
with and without the $\nu$-process, respectively.
}
\label{fig_nupromf}
\end{figure}

 So far we have not mentioned the effects of neutrino process
($\nu$-process), but the process is important for some elements such as 
Li, B, F, and Mn\cite{WH90,YT04,YK05,YK06a,YK06b,YUN08,YS08,SH09,NY10,KI11,IUY12}.
Of course the neutrino emission depends on the explosion model
and thus we have to assume a model to include the effects
(see Ref.~\citen{YUN08} for the detail). 
Fig.~\ref{fig_nupromf} shows the mass fraction distributions of
Li, B, F, Sc, V, and Mn of the supernova ejecta evolved from a 15 $M_\odot$
star with $Z = 0.02$.
In this calculation, we assume that the neutrino luminosity decreases
exponentially with the decay time of 3 s.
The total neutrino energy is set to be $3 \times 10^{53}$ erg.
The neutrino spectra are assumed to obey Fermi distribution with zero
chemical potential.
The temperatures of $\nu_{\mu,\tau}$, $\bar{\nu}_{\mu,\tau}$ and $\nu_e$,
$\bar{\nu}_e$ are set to be 6 MeV/$k$ and 4 MeV/$k$, respectively,
as in Ref.~\citen{YUN08}.
Almost all Li and B are produced through the $\nu$-process in the O/Ne-layer
and He-layer.
When the $\nu$-process is not taken into account, the mass fractions of 
Li and B are smaller than the lowest value in this figure.
Large F production through the $\nu$-process is obtained
in the O/Ne layer (see also Refs.~\citen{KI11,IUY12}).
A part of F is produced in the explosive He burning from $^{15}$N.
Although the F production from $^{15}$N strongly depends on the metallicity,
the F production through the $\nu$-process scarcely depends on the 
metallicity and, thus, it is important throughout the Galactic chemical 
evolution.
Parts of Sc, V, and Mn are produced though the $\nu$-process, especially
in the complete and incomplete Si-burning regions.

\begin{figure}[t]
\centerline{
\includegraphics[width=8cm,angle=270]{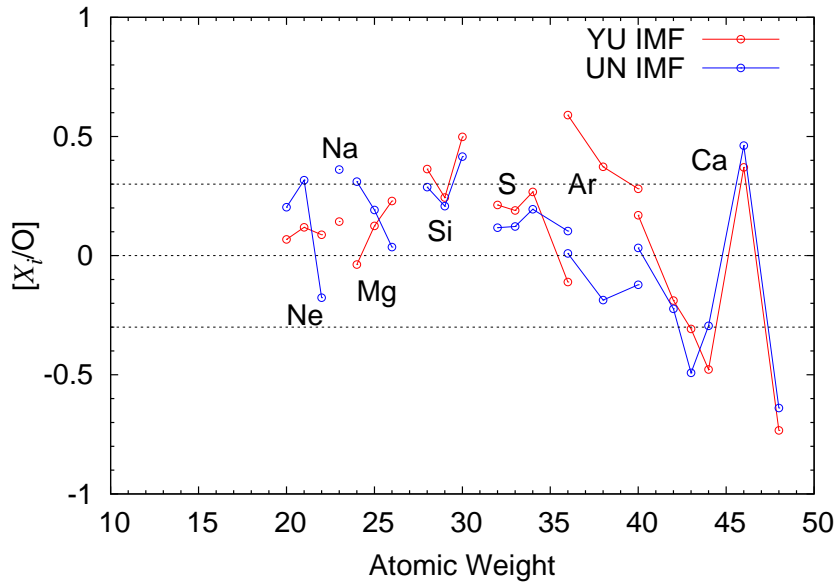}
}
\caption{The IMF weighted yield ratios [X$_i$/O] as a function of atomic weight.
Red and blue lines indicate the yield ratios in YU model and UN model, 
respectively.
}
\label{fig_imf}
\end{figure}

\subsection{\x12c and abundance pattern of Ne to Ca}

 In the previous section, we mentioned that the carbon abundance
after helium burning is important for the abundance of Ne to Ca.
Here we look the results of UN and YU models more closely.
In Fig. \ref{fig_imf} we show [$X_i$/O] vs. atomic number. 
Here $X_i$ represents Salpeter's initial mass function (IMF) weighted
yields of Ne to Ca isotopes, and 
[A/B]= $\log_{10}(Y_{\rm A}/Y_{\rm B})-\log_{10}(Y_{\rm A}/Y_{\rm B})_\odot$,
where $Y_{\rm A}$ and $Y_{\rm B}$ are the abundances of elements A and B. 
If any point is far from the solar value, [$X_i$/O]=0,
the model fails to explain the present day abundance, though chemical evolution
model should be applied for a detailed discussion. This figure show that
for both models, each point is roughly in the $\pm$ 0.3 range, thus 
these models would yield roughly solar abundance pattern.

 Since UN model have larger \x12c after helium burning, this model
has larger Ne/O and Na/O ratios than the YU model as expected.
Larger \x12c means that smaller $X_{\rm C}$($^{16}$O) and thus explosive 
oxygen burning
products, such as S, Ar, Ca, would be less abundant. This is also seen
in the figure. 

The UN yields calculated with the same parameter
choices was applied to the Galactic chemical evolution model in
Ref.~\citen{KU06}. 
It was shown that the model could reproduce the
observed abundance pattern reasonably well. If we look the results closely, 
however,
[Na/Fe] is slightly larger than zero at [Fe/H]=0, and [Ca/Fe] and [Ar/Fe] are
slightly smaller than zero at [Fe/H]=0. This suggests that slightly
larger \cag rate may reproduce the solar abundance better at [Fe/H]=0.
Using the YU codes we will investigate the best choice of the 
\cag rate, or the \x12c, in a forthcoming paper.

\subsection{Zn abundance in EMP stars and asphericity of hypernovae}

 Observations of extremely metal poor (EMP) 
stars, which are defined as halo stars with [Fe/H] $< -4$,  
revealed that [Zn/Fe] and [Co/Fe] are larger than higher metallicity 
stars (e.g., Ref.~\citen{Cayrel04}).
Ref.~\citen{UN02} proposed that these large ratios are explained by the ejecta of
complete Si burning of hypernova. This idea explains also why such EMP
stars have low [Fe/H]. In the early universe where the metal abundance in 
inter stellar matter (ISM) was low, the Fe to H ratio was determined by the amount of Fe produced by a 
supernova over that of hydrogen swept by the supernova shock. Since the swept
hydrogen mass is roughly proportional to the explosion energy, the ejecta
of a hypernova tends to have lower [Fe/H] than that of a supernova.

 As mentioned above, however, there is one
problem for this idea. Zn and Co are produced mainly in the 
complete Si burning while Fe or $^{56}$Ni is produced both in complete and
incomplete Si burning regions. Therefore, [Zn, Co/Fe] becomes larger when
relatively more material is ejected from the complete Si-burning region.
Under the assumption of spherical symmetric explosion, this means that
the mass-cut has to be taken sufficiently deep, then the ejected
mass of Fe or $^{56}$Ni also increases. 
On the other hand if too much Fe is ejected O/Fe ratio, for example,
becomes too small compared with observations. 
To quantify this problem Ref.~\citen{UN02} introduced the
'mixing-fallback model'. In the model, yields are calculated as follow.
First, inner most matter is mixed between the mass coordinate $M_{\rm in}$ 
and $M_{\rm out}$.
For hypernova models $M_{\rm out}$ is set to the upper boundary of the
incomplete Si burning region, and $M_{\rm in}$
is chosen sufficiently deep to
eject Zn. Then it is assumed that only the fraction $f$ of the matter is ejected,
or the fraction $1-f$ fallbacks, from this region.  
To explain the abundance pattern of EMP stars $f \sim 0.1$ is required.

 For supernova models with $M < 25 M_\odot$, $f$=1 is fine to
reproduce abundance pattern of very metal poor stars with
$-3 <$ [Fe/H] $< -2$\cite{TUN07}. This suggests that the
explosion mechanism of supernova and hypernova is different.

 Interestingly almost the same yield as the $f \sim 0.1$ hypernova yield 
could be obtained when we calculate
jet-like hypernova models in 2D\cite{TMU07}. In such models, it is assumed
that an unknown central engine ejects jets along the polar directions.
These jets can blow up the entire star above the Si layer, but the
explosive Si burning mainly takes place along the jet directions. 
As a result the mass fraction of complete Si burning products becomes
smaller than the spherically symmetric models. 
In this paper also in \S 6.3  we show such a 2D calculations applied 
to an initially $M = 110 M_\odot$ star.
Since observed hypernovae show some asphericity\cite{Maeda08}, 
this jet-like explosion model is certainly interesting, 
though we need to know the mechanism for the central engine. 

 It is sometimes claimed that Zn and Co in EMP stars may be explained by the
innermost matter from 'hot bubble' region of normal supernovae 
(e.g., Ref.~\citen{HW10}).
Ref.~\citen{IU10} explored this possibility and discussed that 
this explanation is not likely because fine tuning of $Y_e (= 0.500-0.501)$ 
is required for a 0.06 $M_\odot$ ejecta 
to produce sufficient amount of Zn and Co. 
These conditions are difficult to satisfy
for the hot bubble matter of supernovae, while it is relatively easy 
for the innermost ejecta of hypernovae.

\subsection{Nucleosynthesis of (weak) r-process elements} 
 
  It is quite likely that during explosion supernovae produce elements heavier
than Zn, because some EMP stars have r- and weak-r process elements.
For example CS22892-052
shows r-process abundance pattern which is roughly same as
the present-day r-process pattern.\cite{SN03} 
This suggests that the r-process nucleosynthesis
is almost {\it universal}. There are other classes of stars which show more abundance
of weak r-process elements such as Sr, Y, Zr than the universal r-process pattern.
There are also some stars having enhancement of Mo and Ru as well as Sr, Y, Zr
stars.\cite{HON07} 

 As mentioned above spherically symmetric instant energy injection models cannot
synthesize such elements because to synthesize such elements much larger entropy
and/or more
neutron rich environment are required. On the other hand Fe-peak elements
including Zn can be synthesized from the matter with $Y_e \simeq$ 0.50. 
Most recent 1D core collapse supernovae simulations have shown that the innermost 
matters in the ejecta
have $Y_e \simeq 0.50$ or even greater than 0.5 because of the interaction
with neutrinos (e.g., Refs.~\citen{BRJK06},\citen{MJ09},\citen{F10}).

 Ref.~\citen{IUT09} and \citen{IU10} explored the conditions for synthesizing
weak r-process elements, Sr, Y, and Zr. Typically innermost layers of normal
supernova explosions have entropy per baryon $s/k \sim 5$ and those of hypernovae
have $s/k \sim 15$. With such entropy weak r-process elements are not produced
if $Y_e \simeq 0.50$. 
Thus they relaxed the constraint $Y_e \simeq 0.50$  and considered
$Y_e$ as low as 0.45. Such low $Y_e$ matter cannot be ejected as long as we consider
exact spherically symmetric explosion. However, as shown 
in Ref.\citen{JBR03} 
in 2D calculations inner materials are mixed by convection during explosion
and small amount of low $Y_e \sim 0.45$ and higher entropy $s/k \sim 40-50$
matter, may be ejected. Ref.~\citen{IUT09} showed that
$s/k \gsim 15$ model could produce Sr, Y, Zr explosively. Therefore,
such low  $Y_e$ matter ejected due to the multi-dimensional effects can explain the
origin of weak r-process elements.

 We note that $s/k \lsim 50$ models could produce up to Zr but not heavier elements. 
Because of the lack of observations it is not clear yet weak r- stars which have 
enhancement of
Sr, Y, Zr always have Nb-Mo enhancements as well. If this is the case, innermost
matter of hypernovae or hot bubble of normal supernovae considered in 
Ref.~\citen{IUT09}
may not be the main site for the weak r-process synthesis. 
To produce Nb-Mo much larger entropy ($s/k \sim 150$) is required\cite{IU10}. 
It is currently not clear which astronomical site has such an environment.

\section{Remnant neutron star mass}

 In this section we present remnant neutron star mass in both the UN and YU 
models and discuss their implications. There are several factors to
determine the mass such as the CO-core mass, explosion energy and EOS.
Among them the most important factor is the CO-core mass,
which is determined by the stellar evolution before core-collapse.
Larger CO-core leads larger remnant mass because it typically leads
larger Fe-core and, more importantly it increases the amount of mass
above the Fe-core. 

 Once pre-explosion density structure is given, explosion energy
determines the mass-cut or remnant neutron star mass $M_{\rm rem}$. 
For a given progenitor model,
larger explosion energy blows up more materials above the Fe-core
leading smaller mass-cut.
In this paper, however, as well as our
previous works, we do not determine mass-cut
in this dynamical way but determine it by the amount of 
the ejected $^{56}$Ni amount, $M(^{56}$Ni). This is because $M(^{56}$Ni) is
rather sensitive to explosion energy
when we determine the mass-cut dynamically.
In reality each SN may eject somewhat different amount of $^{56}$Ni. 
However, when we apply the supernova yields for example 
to the Galactic chemical evolution it is more useful to provide
averaged yields. For this purpose it is better to determine the
mass-cut by $M(^{56}$Ni).  

  We set the
explosion energy as $E_{\rm exp}= 10^{51}$ erg, and the ejected $^{56}$Ni 
amount as 0.07 $M_\odot$ to determine $M_{\rm rem}$. 
$M(^{56}$Ni)=0.07 $M_\odot$ is the value for the SN1987A \cite{SNH88}
and considered typical value for normal core-collapse SNe.
Strictly speaking to eject the same amount of
$^{56}$Ni more massive core requires larger $E_{\rm exp}$. 
However, the resultant $M_{\rm rem}$
is not much different, so we fix the $E_{\rm exp}$ 
for simplicity. 

In Table II, we show the baryon mass of the remnant neutron star
mass by $M_{\rm rem}$. 
It correlates with the CO-core mass, which in turn
correlates with the He-core mass.  The UN model leaves smaller 
remnant than the YU model
for the same initial mass, $M$, mainly because \x12c after
He-burning is larger and thus $M_{\rm CO}$ is smaller
as described in \S 3.

As described, the choice $M(^{56}$Ni)=0.07 $M_\odot$ is a typical value,
but this is of course
not the unique vale (see e.g., a review in Ref.~\citen{NT10})
Therefore, we should remind this fact when we compare
with observations. For example, when $M(^{56}$Ni)=0.01 $M_\odot$ 
in the YU model, $M_{\rm rem}$ = (1.38, 1.41, 1.53, 1.53, .172, 1.86, 2.04) 
$M_\odot$
for $M$ = (10, 11, 12, 13, 15, 18, 20) $M_\odot$, respectively.

 The $M_{\rm rem}$ is the baryon mass and is not the observable neutron star mass.
The observed mass is about 10 percent smaller due to the general
relativistic effect and called gravitational mass 
(see e.g., Ref.~\citen{NT87}).
The conversion
from baryon to gravitational mass depends on the EOS of nuclear
matter. In Table II we show the gravitational mass, $M_{\rm g}$, 
for the UU model in Ref.~\citen{UTN94} using 
UV14+UVII EOS\cite{Wiringa88} for nuclear matter.

\subsection{Comparison with the observed mass}

 Here we compare the $M_{\rm g}$ 
of our models with observed neutron star (NS)
masses. NS masses are most accurately determined 
when they are in double NS systems. In this case
if two or more post-Keplerian parameters are obtained, 
NS masses are precisely determined (e.g., Ref.~\citen{Finn94,TC99}).
In Table III we show the NS masses and 
errors for these stars. Interestingly the mass distribution has
peak at 1.33 $M_\odot$ with a small dispersion 0.06 $M_\odot$.
These stars may be considered to keep their birth mass\cite{Ozel12},
so they are suitable to compare with our results.
NSs in binaries with high mass companions are also considered
to roughly keep their birth mass. Ref.\cite{Ozel12} discussed that
the most likely values of the central mass and dispersions
for these NSs are 1.28 $M_\odot$ and 0.24 $M_\odot$, respectively.

\begin{wraptable}{r}{\halftext}
\caption{NS masses in double NS systems and errors.
See references for Ref.~\citen{Ozel12} and therein.
}
\label{table:4}
\begin{center}
\begin{tabular}{ccc} \hline \hline
Name & Mass ($M_\odot$) & Error ($M_\odot$) \\ \hline
J0737-3039 & 1.3381 & 0.0007 \\
 ~pulsar B & 1.2489 & 0.0007 \\

B1534+12   & 1.3332 & 0.0010 \\
 ~companion & 1.3452 & 0.0010 \\

J1756-2251 & 1.40 &  0.02 \\
 ~companion & 1.18 &  0.02 \\

J1906+0746 & 1.248 & 0.018 \\
 ~companion & 1.365 & 0.018 \\

B1913+16  & 1.4398 & 0.002 \\
 ~companion & 1.3886 & 0.002 \\

B2127+11C  & 1.358  & 0.010 \\
 ~companion & 1.354 & 0.010 \\
\hline
\end{tabular}
\end{center}
\end{wraptable}

%
%
%
%
%

 First we note that lowest NS mass in the table is 
$M_{\rm g} = 1.16-1.20 M_\odot$,
and this is consistent with the lowest $M_{\rm g}$ 
model in the YU model $M = 10M_\odot$.
As described above this model may roughly correspond to
the lightest Fe-core collapse supernovae. For a smaller mass 
star such as $M$ = 9.5 $M_\odot$ or less, the star once forms a O-Ne degenerate
core and its collapse may lead a weak supernova explosion\cite{Kitaura06}.

 In the calculation by Ref.~\citen{Kitaura06}, 
the remnant neutron star has $M_{\rm rem}=1.36 M_\odot$ or 
$M_{\rm g} \simeq 1.23 M_\odot$. This seems
to be larger than the observed minimum mass NS,  
though these O-Ne supernovae have not been studied well and currently
only one progenitor model exists\cite{Nomoto83,Nomoto84}, therefore
we do not know the general properties of these SNe yet.

 For the observed double NS typical mass range is 
$M_{\rm g} = 1.27-1.39 M_\odot$
as mentioned above. This corresponds to $M \sim 11.5 - 20 M_\odot$ for
UN model and $M \sim 12 - 14 M_\odot$ for YU model.

 Largest NS mass in the table is $M_{\rm g} = 1.44 M_\odot$ and 
this corresponds to
$M \simeq 22 M_\odot$ for the UN model, and $M \simeq 15 M_\odot$ 
for the YU model.
It is usually said that the border of NS and blackhole formation
is at around $M = 20-25 M_\odot$. Therefore it appears that the $M_{\rm g}$ 
range of the
UN model is consistent with the double NS mass range, though we have not
confirmed yet if the UN model with $M \lsim 11$\msun would result in
$M_{\rm g} \sim 1.2 M_\odot$ NSs. 

 If we adopt the YU model, the mass range of double NS corresponds 
to rather narrow range, $M \sim 12 - 14 M_\odot$, and it is difficult
to understand the reason why. However at this moment we cannot say
that the UN model represents the reality better, because the
number of the double NS systems are still limited. In Ref.~\citen{Ozel12}
using the model of Ref.~\citen{Timmes96} they also discussed that
the narrow mass range of $M_{\rm g}$ 
in double NS systems are difficult
to understand and suggested a particular and rare formation channel
for the systems.
On the other hand, the NS mass in binaries with high mass
companion ranges $M_{\rm g} = 1.04 - 1.52 M_\odot$\cite{Ozel12}.
Although this estimate is more uncertain than that in double NS systems,
this wider range would be more easily understood in the YU model.

 In summary NS mass observations would certainly provide interesting
and important information to constrain the progenitor model, \x12c
and \cag rate. Interestingly all the observed NSs mentioned above
have relatively small mass compared with the ``recycled'' NSs. 
NSs with white dwarf companions, millisecond pulsars and in low-mass
X-ray binaries are called recycled and currently undergoing accretion.
It is now known that these NS can surely be as massive as 
$M_{\rm g} = 2.0 M_\odot$ as the case for J1614-2230.\cite{Demo10} 
Observations and theory are consistent
in the sense that such massive NSs are rarely formed by the
normal supernova explosions. Theoretically, when we fix the explosion
energy, $M_{\rm rem}$ increases rapidly around $M \sim 25 M_\odot$.
This explains why massive NSs are rarely formed.
It will be very interesting to 
constrain observationally that how massive NS can be formed
at birth.

\section{Other recent works of our group}

 In this paper we have described our recent work on stellar
evolution with some new results. Here we briefly review other
works of our group not mentioned above.

\subsection{(Magneto-)Hydrodynamical Simulations}
 
 T. Kuroda and H.U. have developed a 3D magnet-hydrodynamical
general relativistic code with adaptive mesh refinements\cite{KU10}.
This code was used to follow gravitational
Fe core collapse and to calculate spectra of gravitational waves.
We will apply this code to various progenitor models to explore
explosion mechanism and nucleosynthesis.

\subsection{Special Relativistic Hydrodynamical Simulations for GRB Jets}
 
 S. Okita and H.U. have developed a 2D special relativistic
hydrodynamical code. Using this code,
Okita and H.U.\cite{OU12} 
explored the conditions for successful ejection of 
ultra-relativistic jets in the collapsar model of GRBs. 
This code is also applied to explore various aspects of
explosions and nucleosynthesis in SNe and GRBs. One example
is given in the next subsection. 

\subsection{Evolution and nucleosynthesis in very massive stars which 
end up Fe core-collapse SNe}

 Motivated by the discoveries of very luminous SNe,
such as SN 1999as (SN Ic) \cite{Deng01} and SN 2006gy (SN IIn) \cite{Ofek07}, 
H.U. and K. Nomoto\cite{UN08}
calculated evolution of metal-poor massive stars in the mass range 
$M = 20 - 100 M_\odot$ with metallicity $Z=10^{-4}$ to study how much 
$^{56}$Ni is produced in core collapse SNe (CCSNe). 
They found that $^{56}$Ni of $\sim 13 M_\odot$ can be produced for 
the 100 $M_\odot$ star. 
This amount is sufficiently large to explain 
SN 1999as if the SN shines by the $^{56}$Ni decay.

 Since the actual bright SNe Ic usually appears in a host galaxy
having metallicity larger than $Z=10^{-4}$, mass-loss effect becomes
important. Therefore, T. Y. \& H.U.\cite{YU11} studied 
the uncertainties in mass-loss in detail for $Z=0.004$ 
which corresponds to the host galaxy of SN 2007bi. 
SN 2007bi was a very bright SN Ic and $3.6 - 7.4 M_\odot$ of $^{56}$Ni 
is required if it shines by radioactive $^{56}$Ni. 
First Gal-Yam et al.\cite{Gal-Yam09} discussed  that the SN was a 
pair-instability SN (PISN)
but Moriya et al.\cite{Moriya10} showed that they can be a CCSN 
if a $\sim 43 M_\odot$
CO star explodes with $E_{{\rm exp}} = 3 \times 10^{52}$ erg. 

 Since PISN model is much more massive than the CCSN model,
these models predicts quite different light curves.
Unfortunately without the LC data well before the maximum light
one could not distinguish these two models. 
T.Y. and H.U.\cite{YU11} found that for $Z=0.004$, required ZAMS ranges are 
$M = 110-280 M_\odot$ and $M = 515-575 M_\odot$ for
CCSN and PISN model, respectively. 
They provided that, if the progenitor was a single star and assuming the 
Salpeter's IMF, the ratio of the probabilities of CCSN to that of PISN 
appropriate for SN 2007bi is 42.
We should remind, though, for the CCSN model we are assuming that
such a large CO star can explode energetically. So far no one
could have shown such a explosion from the first principle calculations.

\begin{figure}[t]
\centerline{
\includegraphics[width=8cm,angle=270]{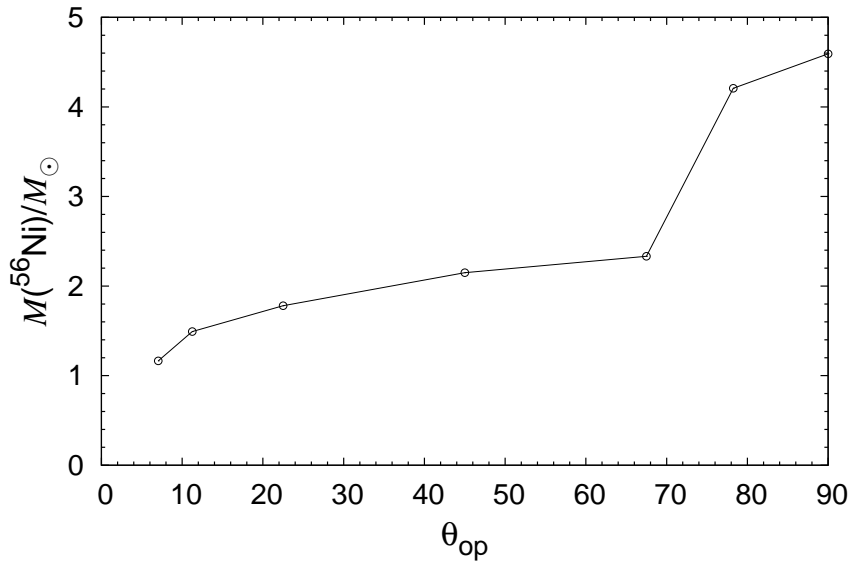}
}
\caption{The relation between the ejected $^{56}$Ni amount 
and opening angle $\theta_{{\rm op}}$ in axisymmetric supernova model.
The progenitor is a 43.2 $M_\odot$ WO star.
The explosion energy is set to be $3 \times 10^{52}$ erg.}
\label{fig_m110axi}
\end{figure}

 Under the assumption that it can explode 
Okita, H. U., and T. Y. simulated spherically symmetric and axisymmetric jet-like
core-collapse supernova explosions of a $M_{{\rm f}} = 43.2 M_\odot$ WO star 
in Ref.~\citen{YU11}.\cite{YU12,OUY12}
This progenitor is evolved from a $M = 110 M_\odot$ star with the metallicity
of $Z = 0.004$.
The CO core mass is $M_{{\rm CO}} = 38.2 M_\odot$.
For spherically symmetric calculations the same method as described in 
\S 4 is used. For the jet-like explosion 
the same code mentioned in \S 6.2 was used. 
After the explosion simulations, nucleosynthesis is calculated
post-processingly.
First we investigate the explosion-energy dependence of the $^{56}$Ni amount.
We found that $^{56}$Ni larger than 3 $M_\odot$, enough to
reproduce the light curve of SN 2007bi, is produced in the supernova ejecta
if the explosion energy is larger than $2 \times 10^{52}$ erg.
The investigation by the jet-like explosions with different opening angles
$\theta_{{\rm op}}$ with $E_{{\rm exp}} = 3 \times 10^{52}$ erg indicated
that the ejected $^{56}$Ni yield strongly depends on the opening angle.
Fig.~\ref{fig_m110axi} shows the relation between the amount of 
the ejected $^{56}$Ni and the opening angle.
Although the spherical explosion releases the 
amount of 4 $M_\odot$ $^{56}$Ni 
in the ejecta, the ejected $^{56}$Ni is much smaller when
the opening angle is smaller than 78$^\circ$.
The ejected amount of $^{56}$Ni is smaller than 2.3 $M_\odot$ for the opening
angle smaller than $68^\circ$.
This suggest that if SN 2007bi was a CCSN,
it exploded with large opening angle or
the explosion was more energetic than $3 \times 10^{52}$ erg.

\subsection{Pop III very massive stars: Evolution, Dark stars, Stability}

  The formation of first generation or Population (Pop) III stars 
in the Universe is considered to 
be quite different from that of later generation stars.
This is because when the first generation stars were formed 
there was no metal yet, thus radiative feedback from the proto star
was weak and might not be able to stop gas accretion. Also gas accretion 
rate itself was larger than the present universe. As a result
it was expected that typical mass of Pop III stars were 
100 $M_\odot$ or more (e.g., Ref.~\citen{YHS03}).

 In Ohkubo, H.U. et al.\cite{Ohkubo06}, they calculated evolution and 
nucleosynthesis in the very massive 500 and 1000 $M_\odot$ stars
assuming that they explode energetically. Next, by taking account of
a realistic mass accretion rate, Ref.~\citen{Ohkubo09} 
calculated Pop III star evolution with the mass accretion.
They found that typically $500 - 1000 M_\odot$ stars were formed. 
It is notable that this mass range is larger than the often said
mass range for a Pop III star, $M = 140-300 M_\odot$. 
If the mass range is in the $140-300 M_\odot$ range, most first stars 
should have exploded as PISNe. 
However, the abundance patterns of EMP and HMP stars
with ${\rm [Fe/H]} < -4$ are not consistent with the yields of PISNe
\cite{UN02}.

 These works, however, neglected the effects on the accretion
disk likely formed around the first stars. 
Ref.~\citen{McKee08} and \citen{Hosokawa11} 
investigated such effects
and showed that the accretion disk was evaporated by the 
radiative feedback. Then the accretion onto a first star
may stop before its mass grows more than 100 $M_\odot$, preventing the
formation of a PISN\cite{Hosokawa11}.

 Even though it is true,
there is still a possibility for the formation of 1000 $M_\odot$ stars.
First stars are formed around the center of a dark halo where 
the density of dark matter is much higher than other places.
Refs.~\citen{SF08,SB09} showed that if the dark matter is 
self-annihilating
WIMPs the self-annihilation energy is sufficiently large to
sustain the first stars. Such stars are called ``dark stars''.
These stars typically have much larger radius, thus surface
temperatures are lower and radiative feedback is expected to be weak.
Hirano, H.U. and N. Yoshida\cite{Hirano11} calculated the
dark star model with mass accretion. We confirmed that the
stars keep large radius until it reaches the main sequence
stage at around $M \sim 1000 M_\odot$. 

 H.U. et al.\cite{Umeda09} also considered
the accreting dark star model including the effects of captured
dark matter
annihilation that becomes important after the main sequence stage.
They showed that if the baryon-dark matter scattering cross section
is as large as $\sigma = 10^{-38}$cm$^{-2}$, 
the dark stars could grow more: it could be
$10^4 to 10^5$ \msun depending on the mass accretion rates.

 It has been known that such huge stars are vibrationally
unstable against epsilon mechanism (e.g., Ref.~\citen{Baraffe01}).
As far as the star is sustained by the DM annihilation and
not nuclear burning, the epsilon mechanism does not operate.
However after the main sequence stage, nuclear burning soon
dominates unless the captured dark matter effect is important.
Sonoi and H.U.\citen{SU12} explored the stability of such very
massive stars against the epsilon mechanism. They found that 
the amount of mass loss is less than 10 percent of the whole
stellar mass.

\subsection{Dust Formation in Supernovae}

 H.U. has been working also on dust formation in supernovae
with T. Nozawa, T. Kozasa and collaborators. 
For example, Nozawa et al.\cite{Nozawa03}
calculated the dust yield for Pop III supernovae including
CCSNe and PISNe, and showed that large amount of dust grains 
would be produced in the early universe by these SNe. 
Ref.~\citen{Nozawa07} calculated the dust destruction 
by the reverse shock in a supernova remnant. The theory of
dust formation was also applied to actual supernovae.
Ref.~\citen{Nozawa08} and \citen{Nozawa10} for 
SN 2006jc and Cas A supernova remnant, respectively, showed that
the observed data are reasonably reproduced by the theory.
It was also applied to SNe Ia\cite{Nozawa11}.
These dust grains in supernovae play important roles in 
the star and galaxy formation, and chemical evolution in
the galaxies.

\subsection{Presolar Grains from Supernovae}

Presolar grains are recovered from primitive meteorites or interplanetary 
dusts and are identified as the grains having very large isotopic anomalies
compared with the solar-system materials 
(e.g., reviews~\citen{Zinner98,Clayton04}).
Observed isotopic ratios of the grains are considered to indicate the 
traces of the nucleosynthesis in stars at their birth or Galactic
chemical evolution.
Small amount of presolar grains are considered to originate from supernovae.
They mainly indicate the excesses of $^{12}$C, $^{15}$N, and $^{28}$Si.
Some supernova grains show evidence for original presence of radioactive 
$^{44}$Ti in Ca isotopic ratios, which strongly supports their origin.
Isotopic ratios of heavy elements such as Mo and Ba have been also observed.
However, it is still difficult to reproduce observed isotopic ratios
by supernova models.
Bulk composition of supernova ejecta of supernova models does not reproduce 
observed isotopic ratios of supernova grains.\cite{Travaglio99}.
In order to reproduce observed isotopic ratios, inhomogeneous mixing is
required.

TY and Hashimoto\cite{YH04} and TY\cite{Y07} investigated supernova
mixtures reproducing several isotopic ratios of supernova originating
SiC and graphite grains.
They divided supernova ejecta into seven different layers and investigated
the mixing ratios reproducing C, N, O, Al, Si, and Ti isotopic ratios
as many as possible for individual grains.
The mixing ratios of the mixtures strongly depend on the reproduced isotopic
ratios.
The main component of the mixtures are the innermost Ni layer and the outer
He/C and He/N layers, so that inhomogeneous mixing of supernova eject
is necessary.
TY, HU, and Nomoto\cite{YU05} investigated the supernova mixtures of different
stellar masses reproducing Si isotopic ratios of supernova grains.
They obtained that less massive supernovae with $M \lsim 15 M_\odot$ and
hypernovae are preferable to reproducing Si isotopic ratios of supernova
grains.

\section{Discussions and Future Prospects}

\subsection{Massive star evolution}

 In this paper we explained the differences in the
newly developed efficient YU code and previous UN code, and shown that the
YU code yields reasonable results as shown in \S 3 in some detail. 
We need such an efficient code because the study of massive star evolution
still requires heavy amount of calculations as described in the followings.

As shown in Tables I and II, the results of UN and YU models
presented in this paper are different mainly because \x12c is different. 
The values of \x12c sensitively depend on the \cag rate
and treatment of convection. Since it is currently not possible to
determine the \cag rate, we need calculate for
various choices of the rate
to find a best set to fit observations. Traditionally
nucleosynthetic argument shown in \S 4.1 has given the most stringent
constraint on \x12c and the \cag rate. However, as shown in Fig.~11,
it is not easy to distinguish even the UN and YU models because 
abundance data are only given for the IMF weighted yields.

 We propose here that NS mass distribution will give
alternative and independent constraints on the \x12c and \cag rate. 
As shown in \S 5 and Table II, UN and YU models predict different
mass distribution for the remnant NSs. Though the differences are
not so large, observed NS masses are given in very fine
precision for the double NS systems (Table IV). Therefore, by calculating
stellar evolution in binary systems, we may
constrain the \x12c and \cag rate more precisely, though it
requires many calculations for various possible binary pairs.

 One of our purposes for developing the efficient code is to
tackle on the evolutions just below and above the Fe-core forming
SNe. As mentioned in \S 3.1, such stars experience violent
shell flashes near the end of their evolution. It is not clear 
yet if such events cause mass loss and shock interactions to be
observable. In order to follow these stages precisely, it is 
important for a code to include the acceleration term. As mentioned
in \S 2, we will include the term into the YU code in near future.

 The stars just below the critical mass for the
Fe-core formation is very interesting for another reason.
Such a star may become an electron capture supernova (ECSN). 
We would like to construct another progenitor models of ECSNe
than the one by Ref.~\citen{Nomoto84} to see if the successful explosion
by Ref.~\citen{Kitaura06} is model independent or not.
Since these calculations require high numerical resolution
in both time and space\cite{Poelarends08}, we need an efficient
code to perform computation. 

 We have been working also to develop a new code including the
rotation effects as in Refs. \citen{HL00,HM04,HWS05,YL05,WH06,EM08}.
This code is based on the YU code and aiming for efficient 
computation as well. Although rotating stellar models in the 1D
formalism have been already calculated by these authors,
angular momentum transfer is still quite uncertain. 
Therefore, the evolution of a rotating star is still far from 
complete understanding. Since rotation is inevitable for
constructing realistic progenitor models for hypernovae and GRBs,
and possibly even for normal CCSNe,
we plan to calculate rotating progenitor models as well.

 The uncertainties in mass-loss rate is the another reason why
one needs to compute several cases to find a better set.
Since the knowledge for the rate is still very limited, one needs 
to constrain the rate by using various information including
the properties of massive stars, supernovae, compact remnants,
ISM abundances and so on. Unfortunately this may not be easy because other
uncertain factors may be involved such as rotation, binarity and
magnetic field effects. Supernovae may provide a key to understand
the rate. For example, as discussed in Ref.~\citen{YU11} if one
can identify SN 2007bi-like event to a CCSN or PISN, we can 
severely constrain the mass loss rates.

\subsection{Nucleosynthesis}

 As for our nucleosynthetic works, in \S 4 we described the
success and limitations of the simple 'instant energy injection'
model. This model reasonably reproduce supernova yields
up to Zn. Although spherically symmetric hypernova models
over-produce Fe to lighter elements ratio, such as Fe/O, this 
problem may be avoided if one consider jet-like explosions in 2D
for hypernovae even under the instant energy injection model.
This suggests that the nucleosynthesis up to Zn does not much depend
on the detail of the central engine. 

 On the other hand, nucleosynthesis of r- and weak r-process
elements depends on the detail of the explosion model.
This in turn implies that successful supernova and/or hypernova and/or
GRB models must produce these elements. 

 As described in Ref.~\citen{IUT09} there are observational
indications that normal CCSNe produce lighter weak r-process
elements, Sr, Y and Zr. These elements can be produced and
ejected from the hot bubble of normal SNe. However,
to produce heavier weak r-process elements, Mo and Ru,
much larger entropy and neutron rich environment is required\cite{IU10}.
It is currently not certain if such an environment can be
realized in a CCSN. This is because long term simulations of a CCSN
showed that high-entropy matter ejected from a CCSN is
always almost neutral or proton-rich.\cite{F10} 
As mentioned in \S 4, it is not
clear if all Sr-Zr rich EMP stars are also Mo-Ru rich.
If this is the case, a CCSN model has to produce Mo-Ru somehow.
One possibility is the $\nu p$-process 
(Refs.~\citen{PW05,Fr06,Wa06,Wa11})
not considered in Ref.\citen{IU10}. With this process, such weak 
r-process elements may be synthesized in proton-rich matter if
entropy is sufficiently high. Thus it is critically
important to observationally clarify if normal SNe produce
Mo-Ru or not. We should note that the results in Ref.~\citen{F10}
are not obtained by the first principle calculations, but 
the explosion is driven by artificially
enhancing the effective neutrino luminosity.
The main reason that the matter becomes proton-rich is 
because of the interaction between matter and neutrino. 
If the explosion is driven by the assist of something else,
such as the rotation energy, neutron rich matter may be ejected.
 
 The arguments above also applies to the r-process elements.
Currently there is no clear evidence in the EMP stars that 
r-process elements are produced in normal CCSNe and
hypernovae. For example, Zn-rich EMP stars, that we consider them
polluted by a hypernova, show no enhancement of r-process elements. 
From the results in Ref.~\citen{F10}, it seems hopeless to
produce r-process elements in normal CCSNe, 
though unknown massive stars should have polluted
metal poor r-rich stars, such as CS22892-052.
Some authors consider NS-NS or NS-blackhole 
merger systems for r-process sites 
(e.g., Refs.~\citen{FR99,Surman08,Metzger09,Goriely11,Caba12,Wana12}) 
though it is not clear yet if such sites
are sufficient to explain the whole r-process abundance 
in the universe. Therefore it is still interesting to explore
various SN explosion models to examine if they can produce
r-process elements or not. 
 




\section*{Acknowledgements}

We thank Hideyuki Saio for providing the stellar evolution code
and useful comments.
We are grateful to K. Nomoto and S. Okita for useful discussions.
We acknowledge support by the grants-in-aid for Scientific Research
(20041005, 20105004) from the MEXT of Japan.

\end{document}